\documentclass[pra,preprint,showpacs,endfloats*,aps]{revtex4}
\usepackage[dvips]{graphicx}
\begin{document}

\title{Structural and angular dependence of coercivity and magnetic 
remanence of electrochemical ferromagnetic nanowires}

\author{A. Ghaddar$^{1}$}\email{abbas.ghaddar@univ-brest.fr}
\author{F. Gloaguen$^{2}$}
\author{J. Gieraltowski$^{1}$}

\affiliation{$^{1}$ Laboratoire de Magn\'{e}tisme de Bretagne, UBO, \\
 CNRS-FRE 3117, C. S. 93837 Brest Cedex 3, France \\
$^{2}$ Laboratoire de Chimie, Electrochimie Mol\'eculaire et Chimie Analytique, UBO, \\
CNRS-UMR 6521, C. S. 93837 Brest Cedex 3, France}

\date{\today}

\begin{abstract}
A novel method for controlling nanowire magnetic properties and growth
from filling time profile is presented.\\
The wires are grown with an electrodeposition method ("Template synthesis")
with a wide selection of pore diameters.
We show that stray-fields presence in ferromagnetic nanowires are
entirely dependent on the nanowire diameter. Besides a crossover effect
in the reversal mechanism is observed with change in diameter.
In this work, theory and experiment agree and confirm that 
according to the variety of hysteresis loop measured, about four ranges of values
of pore diameter control the orientation of nanowire magnetization easy axis 
with respect to the geometrical axis.


\end{abstract}

\maketitle

\section{Introduction}

 There has been an increasing interest~\cite{Zheng} in the fabrication and properties of 
nanostructured magnetic materials not only from a fundamental but also from a 
technological point of view, owing to their potential applications in magnetic 
recording media~\cite{Manalis}, biosensors and magnetic sensors based on the giant 
magneto-resistance effect~\cite{Yoshida, Piraux}. The ultimate density of recording media depends 
on the size of the individual magnetic elements. The miniaturization of the 
latter can now be realized through the electrodeposition of nanodots or 
nanowires into self-assembled arrays, such as track-etched polymer membranes 
and anodic alumina filters.

Structural and magnetic characterization of arrays of
Nickel (Ni) nanowires produced by electrodeposition in polycarbonate membranes
with diameters in the (15-100 nm) range and $L=6  \mu$m average length are made. The
nanowires made within polycarbonate membranes, with various diameters (15 nm,
50 nm, 80 nm, 100 nm), are used to study the influence of the various
parameters such as pore geometry and deposition process on the magnetic
properties. Magnetization curves with various magnetic field orientations
and nanowire diameters were determined at room temperature by magnetometry.
According to the form of hysteresis loop measured, about four ranges of values
of pore diameter control the orientation of easy axis of magnetization with
respect to the wire axis. Reducing the diameter of the nanowires from 100 to
15 nm leads to increasing coercive fields from 347 to 590 Oe.
The measured coercivity as a function of angle ($\varphi$) between the
field and wire axis reveals that the coercive field decreases
(increases) with angle, peaking at $\varphi =0{}^\circ $,
for nanowire diameters smaller (larger) than 50nm. 

 "Template synthesis" is an elegant chemical approach to the manufacturing of 
nanostructured materials, in particular for different kinds of nanowires~\cite{Brumlik}. In 
this paper, arrays of Ni nanowires have been obtained by filling a porous 
polycarbonate membrane, wich contains a large number of cylindical holes with a 
narrow size distribution. Characterization and understanding of the magnetic 
properties of nanowire arrays have been a challenge for years. Some problems 
remain unclear. For example, there are still open questions about the 
mechanisms responsible for the magnetization reversal. The intrinsic propeties 
of nanowire arrays are directly related to the properties of the nanoporous 
template such as the relative pore orientations in the assembly, pore size and 
its distribution, as well as interpore distance.

 Two reversal modes have been suggested as being important: curling and 
coherent rotation. The critical diameter between coherent rotation and curling 
is $d_{c} =2\sqrt{\frac{A}{\pi } }  \frac{q}{M_{S} } $, ($d_{c} $ is also called 
the coherent diameter), $A=1.5\times 10{}^{-6} $ erg/cm is the exchange stiffness 
constant~\cite{Bertotti}, $M_{S} $ the saturation magnetization and q is a solution of a 
Bessel equation~\cite{Aharoni1} (see section 3.1).  

 For wire diameter $d<d_{c} $, the Stoner-Wohlfarth coherent rotation model 
applies~\cite{Stoner1}.

 On the other hand for wire diameter  $d>d_{c}$, the reversal occurs by 
inhomogenous reversal (curling). Increasing further the diameter until $d\gg 
d_{c} $, domains may form within the wire and the magnetization reversal, thus, 
may occur by domain wall motion~\cite{Skomski}.

 As other extrinsic or hysteretic properties, coercivity (or coercive field) 
$H_{C} $, is strongly real-structure dependent. It is one of the most important 
properties of magnetic materials from the viewpoint of their technological 
utilization~\cite{Kronmuller} and the understanding of its intrinsic mechanism is a permanent 
challenge. As different magnetization reversal mechanisms would give a 
different angular dependence of the coercivity $H_{C} $, the measurement of 
$H_{C} (\varphi )$ would provide helpful information about the rotation 
mechanisms. Here, $\varphi $ is defined as the angle between the field 
direction and the wire axis (Fig.1). For anisotropic, perfectly oriented 
polycrystalline magnets based on noninteracting particles, the coercive field 
$H_{C} $ should be equal to the anisotropy field  $H_{A} $. It is known~\cite{Sagawa}, 
however, that the coercive field measured in permanent magnet materials is 
typically one order of magnitude lower than the anisotropy field. It is 
basically accepted that the origin of this discrepancy lies in the distribution 
of local alterations of the magnetic properties due to a variety of 
microstructural features giving rise to magnetization reversal through a 
process wich includes a succession of mechanisms, as well as in the role played 
by interactions, either dipolar and/or exchange, during demagnetization. 

 Aside from coherent rotation, important coercivity mechanisms are, curling, 
localized nucleation, and domain-wall pinning. With increasing size, the 
nucleation mechanism in perfect ellipsoids of revolution changes from coherent 
rotation to curling. However, both coherent rotation and curling greatly 
overestimate the coercivity of most magnetic materials. Nucleation refers to 
the onset of magnetization reversal and determines the coercivity in nearly 
defect-free magnets.

 Most works have concentrated on individual wires~\cite{Lederman, Wernsdorfer}, few~\cite{Bantu, Garcia} reported on 
detailed measurements of $H_{C} (\varphi )$ for nanowire arrays. 

 In this work, we set out to study the magnetic properties of various types of 
nanowires possessing different diameters. Nanowires used in this work are 
fabricated in polycarbonate membranes, with various diameters (15 nm, 50 
nm, 80 nm and 100 nm) in order to study the influence of the parameters 
such as pore shape and deposition process on the magnetic properties.

\begin{figure}[htbp]
\centerline{\includegraphics[width=3.4in,height=2.97in]{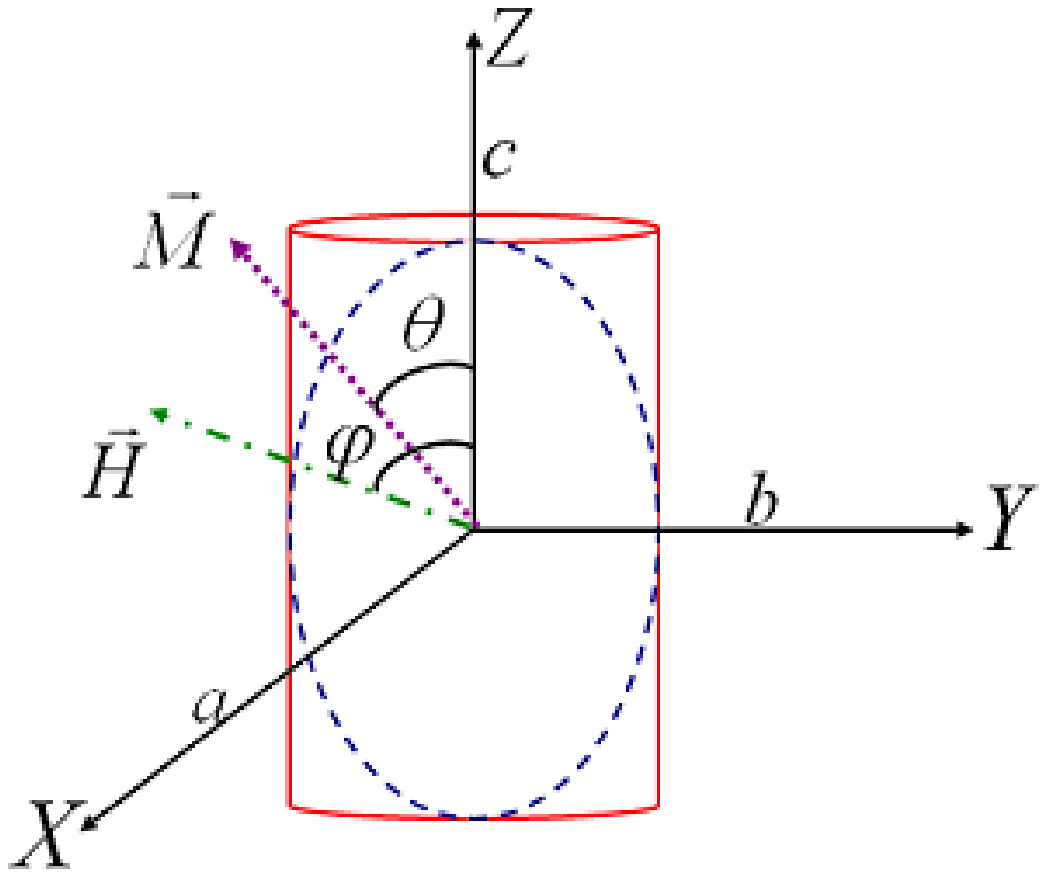}}
  \caption{Single domain particle with uniaxial anisotropy in presence of 
applied field $H$, showing angles, $\theta $ and $\varphi $ that $\vec{M}$ and 
$\vec{H}$ make with wire axis along $Z$, $(a=b<c)$.}
\label{fig1}
\end{figure} 

 We infer from our study that basically four ranges of diameter values are important to 
determine the orientation of the easy axis with respect to the nanowire axis.

\section{Experimental}

\subsection{ Sample preparation}

The nanowire samples are prepared by electrodeposition into the pores of 
nuclear-track-etched polycarbonate membranes PCTE (see Fig. 2) using an 
Autolab Potentiostat employing a three-electrode configuration and an 
electrolyte contaning ion of the metals to be deposited~\cite{Whitney}.
 
\begin{figure}[htbp]
\centerline{\includegraphics[width=3.4in,height=2.97in]{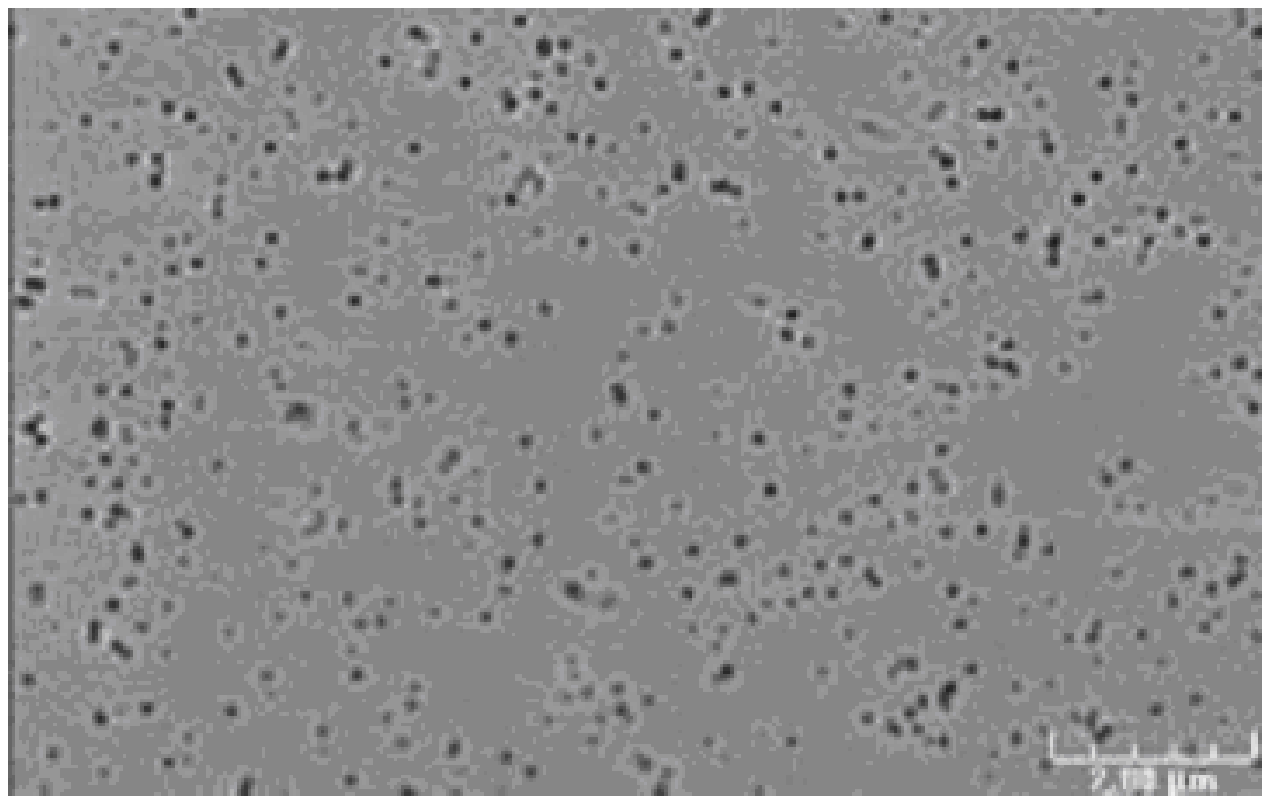}}
  \caption{Polycarbonate PCTE membrane with 80 nm pore diameter.}
\label{fig2}
\end{figure} 

 During sputtering, a Copper layer was sputter-deposited on one side of the 
polycarbonate membrane and used as the working electrode to fabricate an array 
of Ni nanowires. A carbon plate and Ag/AgCl are used as the counter electrode 
and reference electrode, respectively. The work was carried out at a constant 
potential $-1V$ (Ag/AgCl) at room temperature. The following aqueous 
electrolytes used was Nickel sulfate salt (NiSO$_{4} $.H$_{2}$O)(44g/l) in a 
boric acid solution (H$_{3}$BO$_{3}$ )(40g/l). The latter was used as a buffer. 
The study of the magnetic properties of the wire demanded that deposition 
should stop when there is no covering of the upper surface of the membrane by a 
polluting layer of the magnetic material. During the deposition process, the 
time dependence of electrical current was monitored and recorded. Hence, the 
current intensity-time curve recorded during the electrodeposition process 
revealed four different regions (see Fig.3): 

\begin{figure}[htbp]
\centerline{\includegraphics[width=3.4in,height=4in,angle=-90]{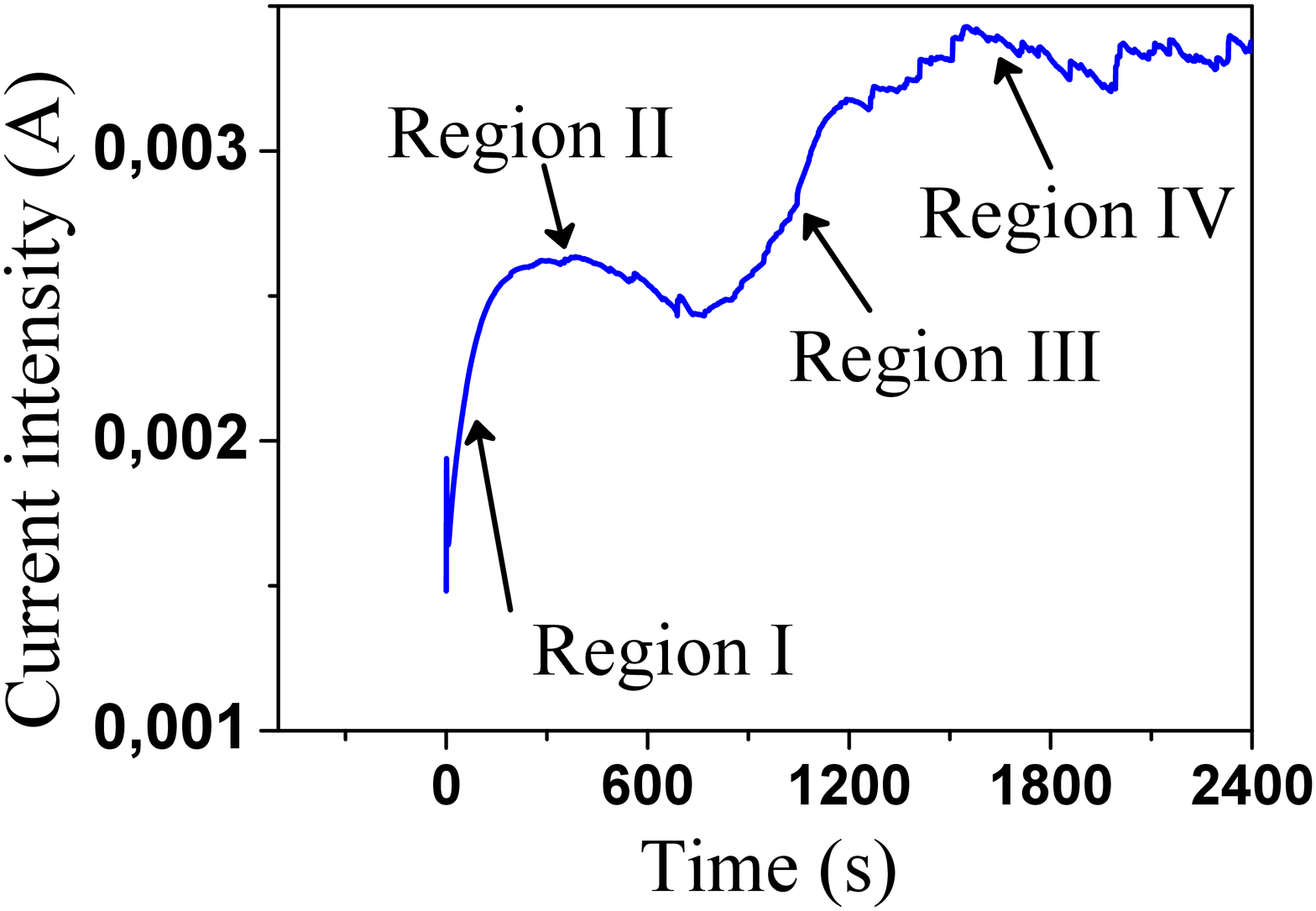}}
  \caption{Typical time variation of cathode current during 
electrodeposition of Ni nanowires arrays or we see the four regions, coverage 
of the pore walls by the ions (region I), filling of the pore interior by the 
growth (region II), fast variation trigged by growth at the upper extremity of 
the pore and subsequent outgrowth beyond the pore. This occurs when the pores 
are completely filled with the material, and the electrodeposited material 
begins to form hemispherical caps over the nanowire ends (region III), 
percolative growth outside the pores (region IV).}
\label{fig3}
\end{figure} 
 
 - Region I: coverage of the pore walls by the ions. \\

 - Region II: filling of the pore interior by the growth. \\

 - Region III:  fast variation triggered by growth at the upper extremity of 
the pore and subsequent outgrowth beyond the pore. This occurs when the pores 
are completely filled with the material, and the electrodeposited material 
begins to form hemispherical caps over the nanowire ends (see Fig. 4). \\

 - Region IV is percolative growth outside the pores. \\

 Attention has been focused on the region labeled II because it is always the 
most complicated to reproduce.

\begin{figure}[htbp]
\centerline{\includegraphics[width=4in,height=3.4in]{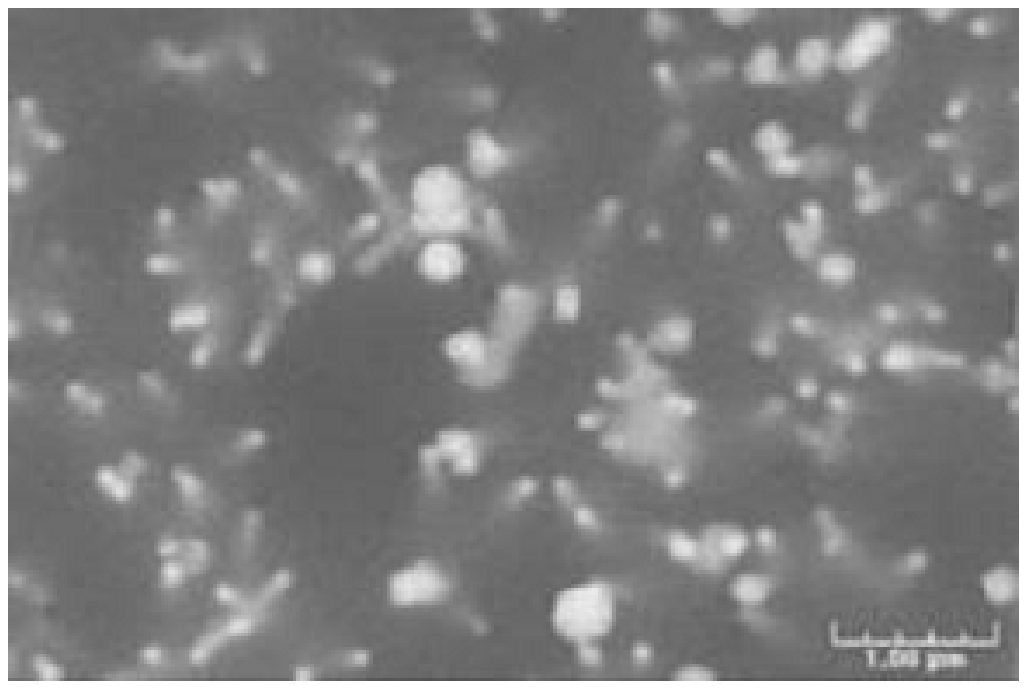}}
  \caption{SEM observation of the top surface where the Ni nanowires 
emerge from the polycarbonate PCTE membranepores in region III.}
\label{fig4}
\end{figure} 

 To understand the mechanism of growth materials in porous membrane, and to 
estimate the time of filling in the pores, a small comparative study between 
two electrodes was made. We tack two Copper electrodes in the same area, which 
was stuck on the first porous membrane (electrode1). According to the nickel 
growth curves on two electrodes (current versus time), we can clearly see the 
difference between the two curves because on the first membrane electrode 
(electrode 1), the deposit is made on small surfaces while on the another takes 
place over the entire surface. After some time, the two curves intersect in a 
single point, which shows that the pores are filled by materials and it is time 
for the emergence of a film on the surface of the membrane (Region IV, Fig.3). 
This also indicates that deposition occurs on two equal surfaces (Fig. 5a). 

 \begin{figure}[htbp]
\centerline{\includegraphics[width=3.4in,height=4in,angle=-90]{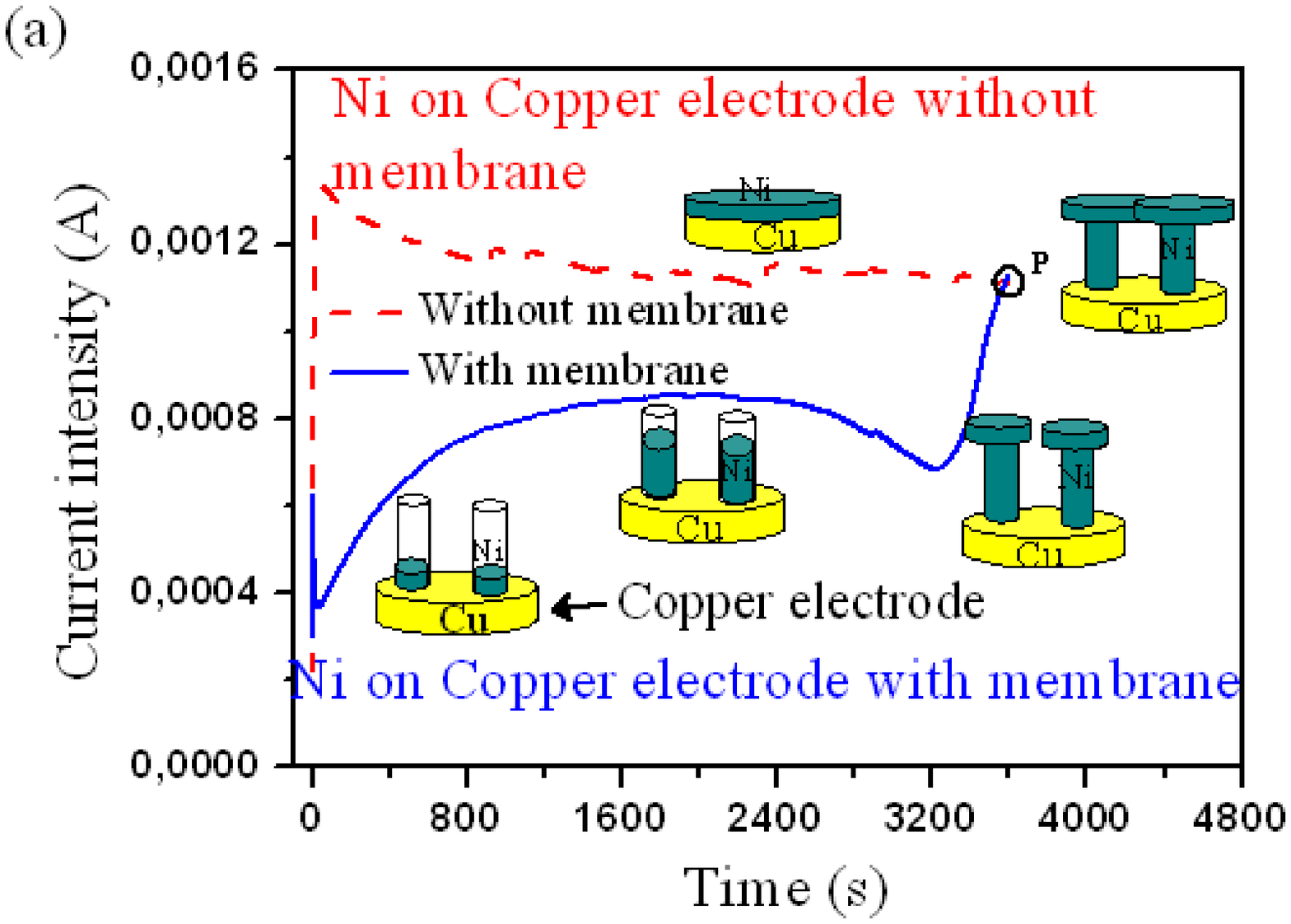}}
\centerline{\includegraphics[width=3.4in,height=4in,angle=-90]{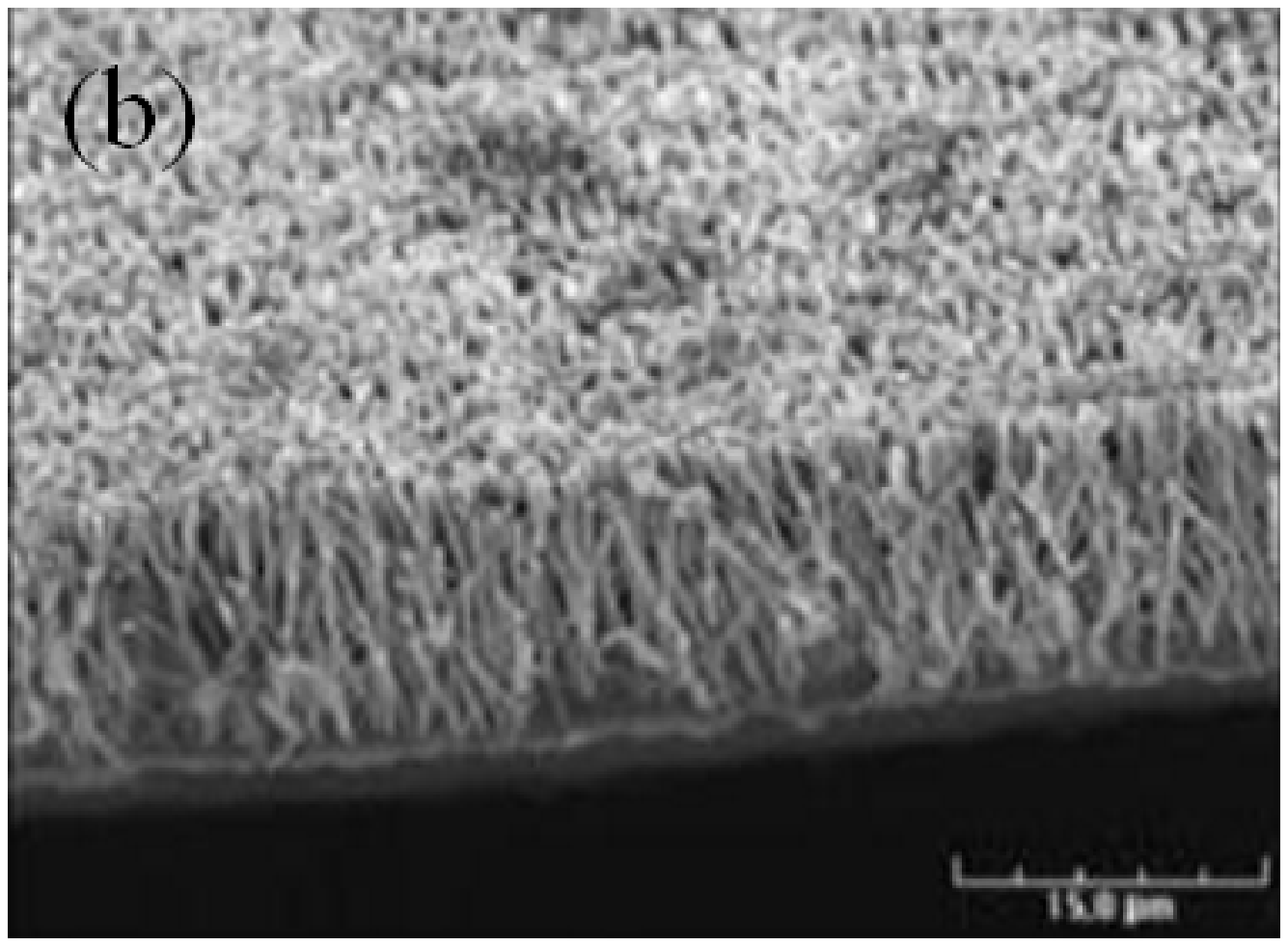}}
  \caption{ Time variation of cathodic current during electrodeposition of 
Ni nanowires arrays when we can determine the exact time of filling of the 
pores (a) the two intensity curvus cut at point P where the vaalus of the top 
surfaces are the same. SEM image of a sample in which all the nanowires are 
misaligned perpendicularly to the polycarbonate membrane plane (b).}
\label{fig5}
\end{figure} 

 Figure 5b shows an image of a sample in which all the nanowires are misaligned 
perpendicularly to the polycarbonate membrane plane. The majority of nickel 
nanowires tilt randomly a few degrees, with respect to the polycarbonate 
membrane plane normal.

\subsection{Magnetization measurements}

 The magnetic properties of the Ni nanowire array are studied by VSM (vibrating 
sample magnetometer), measurements which are performed at room temperature. The 
hysteresis loops at different angles between the magnetic field and the axis of 
the wires were measured.

 Figure 6 shows four cases for the hysteresis loops with the external field $H$, 
perpendicular and parallel to the sample plane (i.e. parallel and perpendicular 
to the axis of the nanowires) for 4 samples with different diameters of (a) 
$15$, (b) $50$, (c) 80 and (d) 100 nm respectively. 

 The hysteresis loops and the coercivities change smoothly with the change in 
the applied field direction.

 \begin{figure}[htbp]
\centerline{\includegraphics[width=3.4in,height=4in,angle=-90]{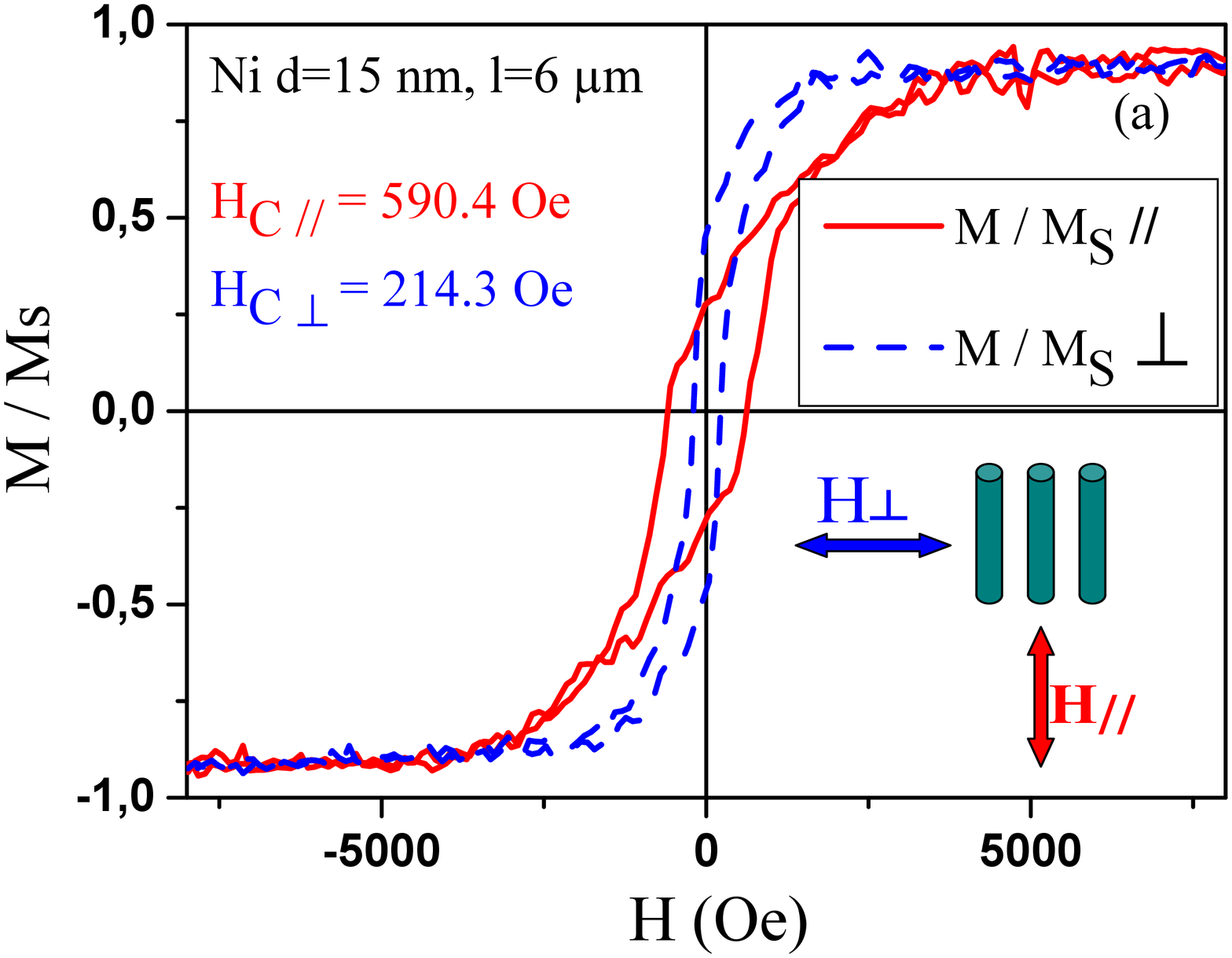}}
\centerline{\includegraphics[width=3.4in,height=4in,angle=-90]{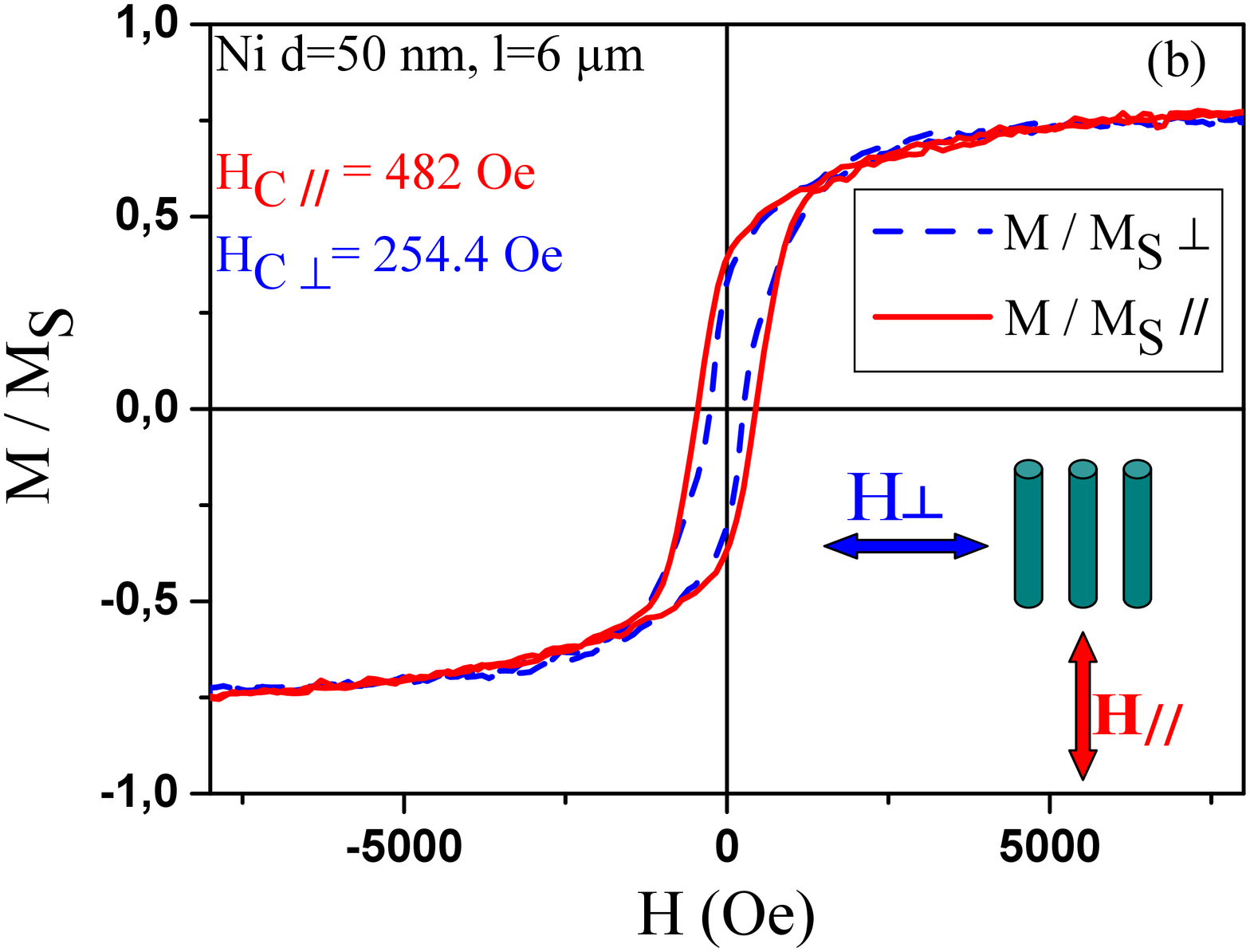}}
\centerline{\includegraphics[width=3.4in,height=4in,angle=-90]{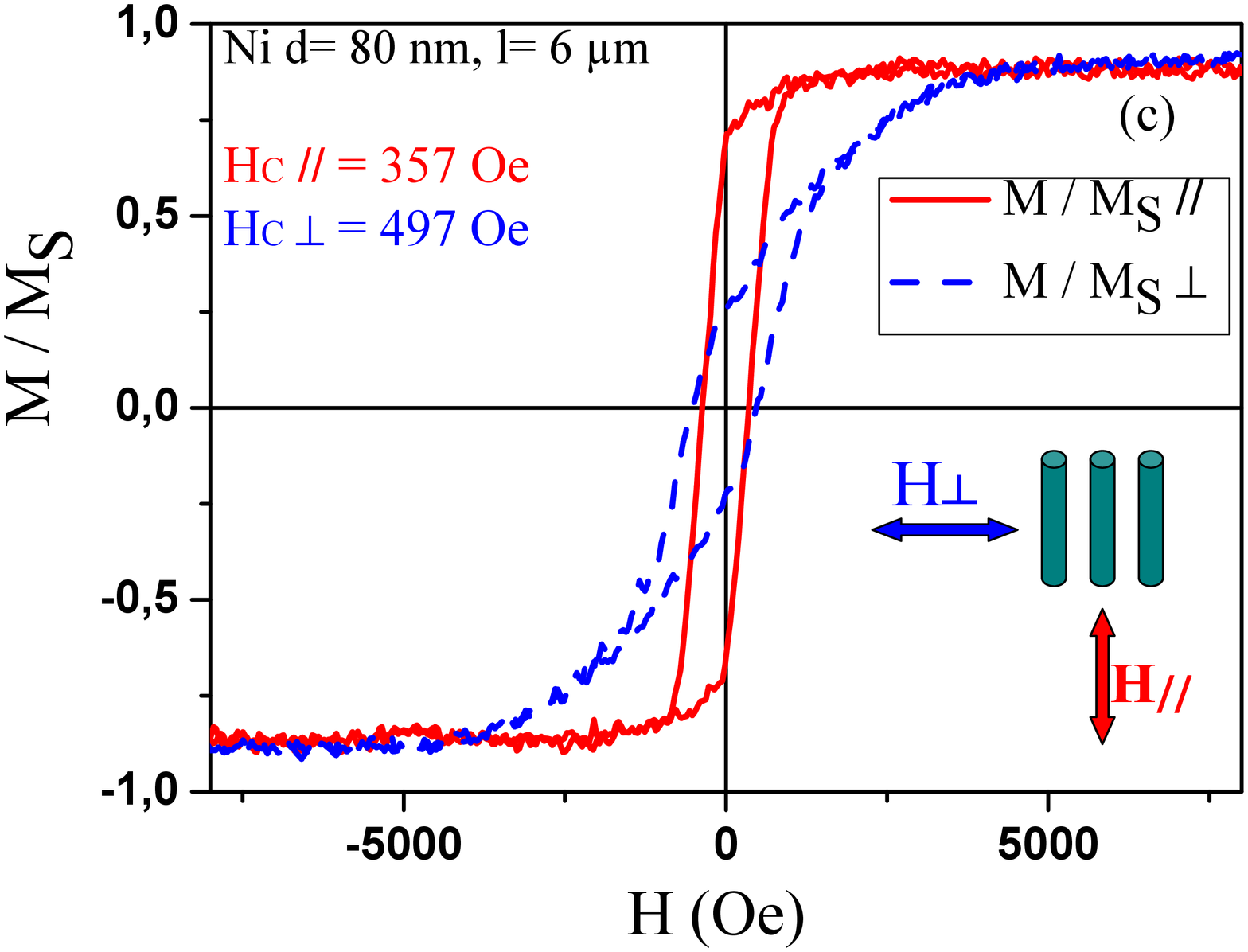}}
\centerline{\includegraphics[width=3.4in,height=4in,angle=-90]{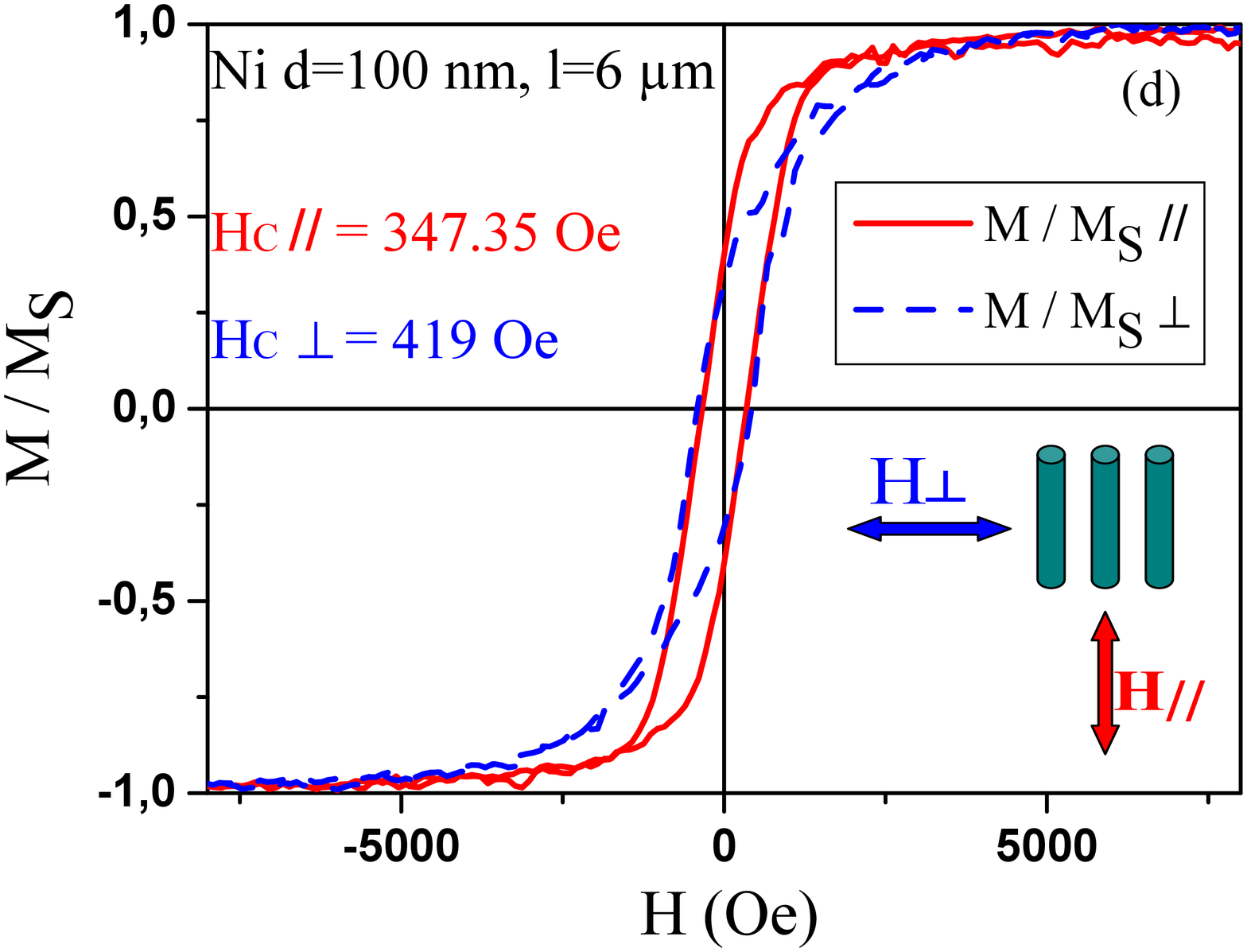}}
  \caption{
a- The hysteresis loops when the external field is perpendicular 
and parallel to the axis wires for 15 nm and $6 \mu$m length.\\
b- The hysteresis loops when the external field is perpendicular 
and parallel to the axis wires for 50 nm and $6 \mu$m length.\\
c- The hysteresis loops when the external field is perpendicular 
and parallel to the axis wires for 80 nm and $6 \mu$m length.\\
d- The hysteresis loops when the external field is perpendicular 
and parallel to the axis wires for 100 nm and $6 \mu$m length.}
\label{fig6}
\end{figure} 

 The saturation field, when the field is applied parallel to the axis of the 
nanowire, is greater than that when the applied field is perpendiculary to the 
nanowire for the 15 nm diameter sample. On the other hand we see the opposite 
case for 80 and 100 nm diameter samples. For the 50 nm sample the parallel and 
perpendicular saturation fields are equal. (See Fig. 6).

 The coercivity dependence on the angles is shown in (Fig. 7a) demonstrating 
that coercivity decreases with the angle for nanowires that have 15 and 50 nm 
in diameter. In contrast, this is not the case of nanowires with 80 and 100 nm 
diameters.

 \begin{figure}[h!]
\centerline{\includegraphics[width=3.4in,height=4in,angle=-90]{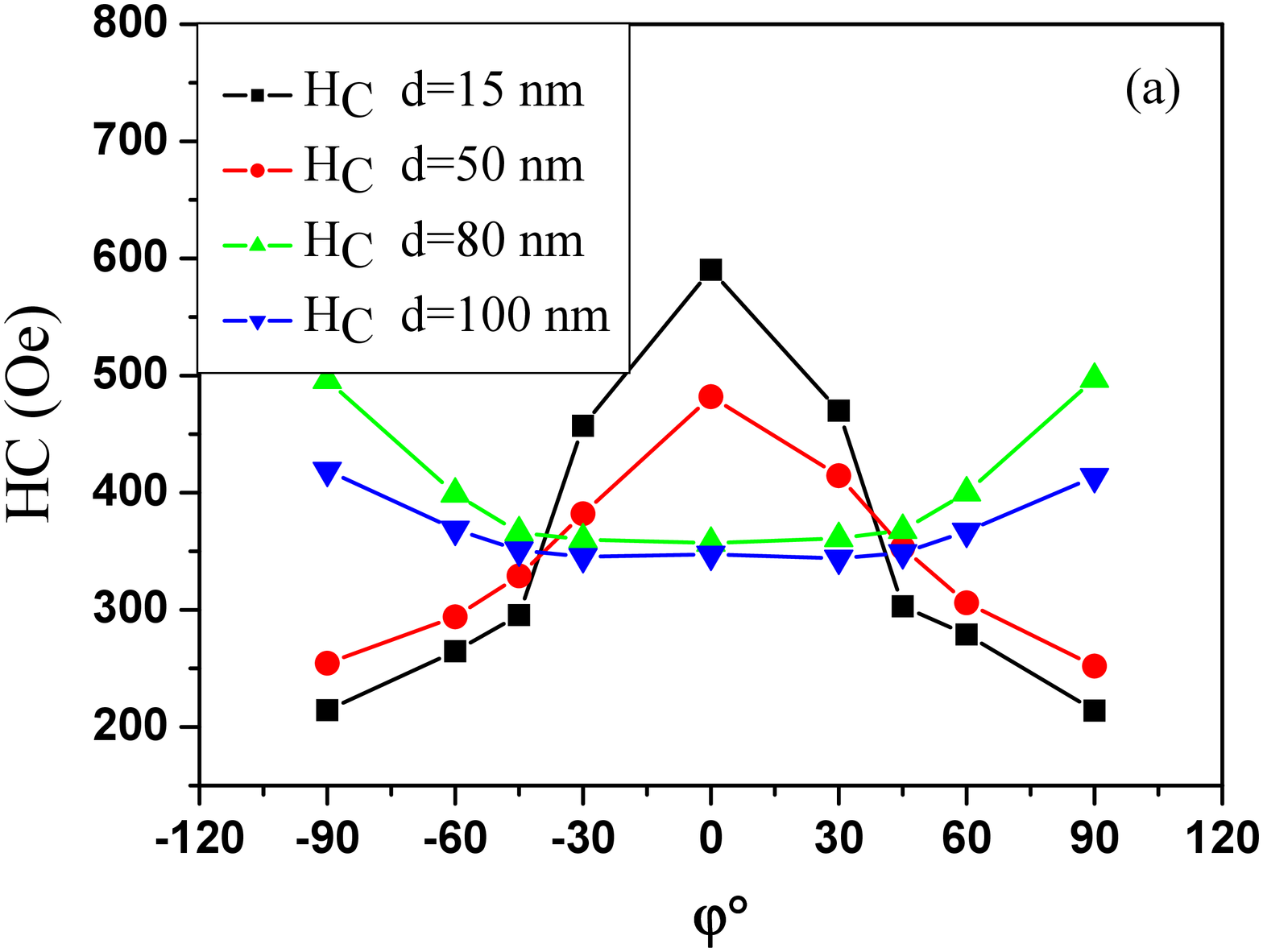}}

\centerline{\includegraphics[width=3.4in,height=4in,angle=-90]{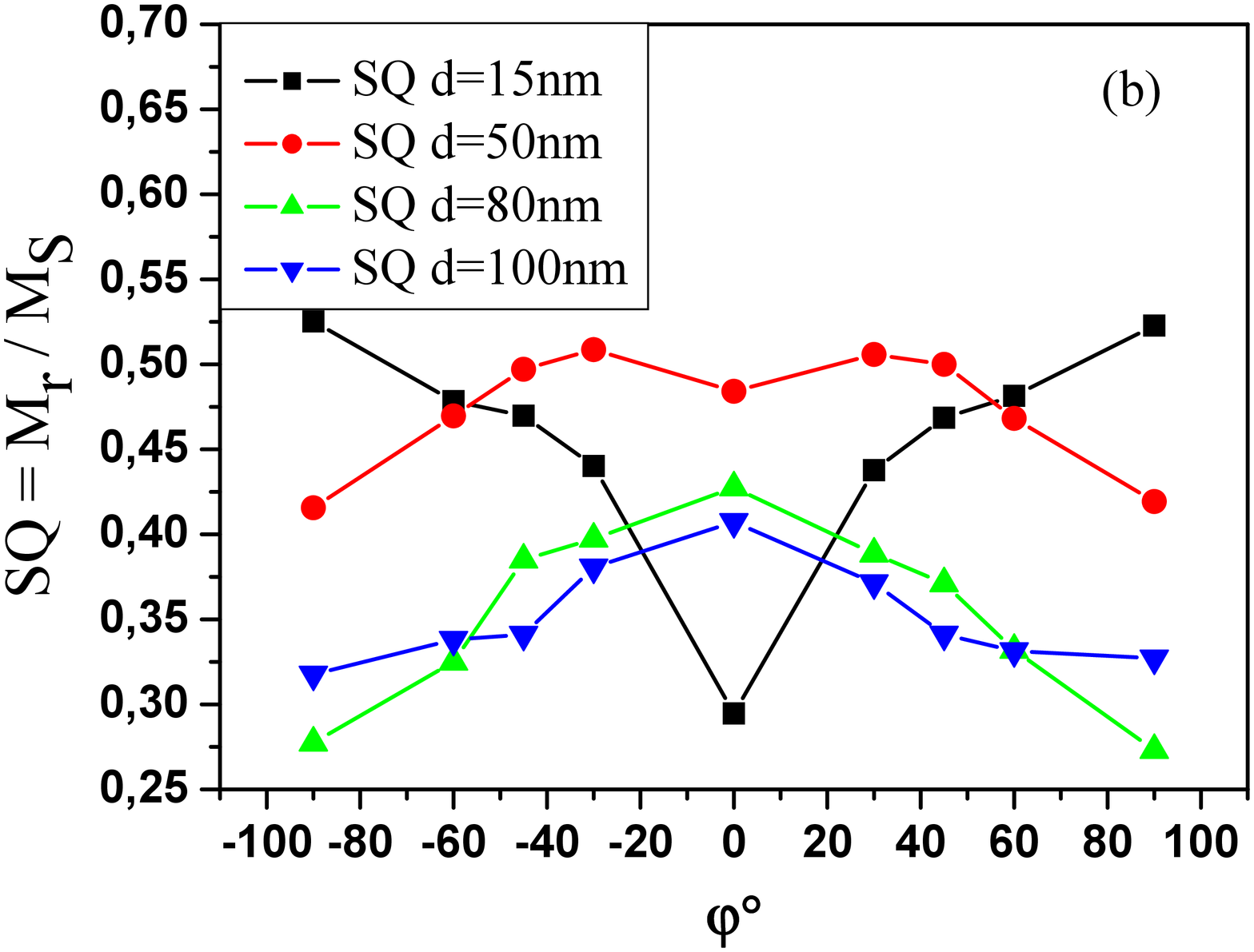}}
  \caption{Angular dependence of the coercivity $H_{C} (\varphi )$ (a) and the squareness $SQ$ (b) of 
nanowires in PCTE membranes with different diameters: 15, 50, 80 and 100 nm 
and $6 \mu$m length.}
\label{fig7}
\end{figure} 

 FIG.7 (a-b) show the angular dependence of $H_{C} (\varphi )$ and 
$SQ(\varphi )=\frac{M_{r} }{M_{S} } $ (squareness) for Ni nanowire arrays in 
PCTE membranes, respectively.

 However, $H_{C} (\varphi )$ shows a quite different behavior for, $d$=15 
nm, 50 nm and $d$=80 nm, 100 nm (Fig. 7a). For $d$=15 nm and 50 nm, we 
observe a bell-type $H_{C} (\varphi )$ variations, but for $d=$80 nm and 100 
nm a bowl shape $H_{C} (\varphi )$ curve is observed.

 These results imply that there is a crossover transition in magnetization 
reversal mechanisms for $d\le $ 50 nm.

 As shown in (Fig. 7b), for the two samples, 80 and 100 nm diameters, the 
squareness $SQ(\varphi )$ of Ni nanowire array decreases with increase of 
angle, peaking at $\varphi =0{}^\circ $, which is the easy axis of the arrays as 
can be seen from the magnetization curves (Fig. 6).

 On the other hand, for the sample of 15 nm diameter the easy axis is 
perpendicular to the wire axis. For the 50 nm diameter sample, easy 
magnetization axis is off to an angle of $30{}^\circ $ of the wire axis.

 \section{Theoretical}

 In this work we can infer that single domain in the nanowires, and 
magnetization reversal modes can be modeled by homogeneous rotation when there 
is a critical size below which a particle remains in a single-domain state 
during switching, or inhomogeneous space dependent reversal $\theta (r)$ when 
the particle size is larger then the critical size, but still in the 
single-domain regime. We assume that all our samples are in the single-domain 
regime~\cite{Skomski}.

\subsection{Magnetization in single-domain particles}

 In principle, the magnetization configuration in a magnetic nanowire can be 
determined from the Brown equations by minimizing the total free energy~\cite{Frei}. In 
bulk ferromagnetic materials, the energy of the system can be minimized by 
forming multiple magnetic domains within which the magnetic moments are 
aligned. However, there is a critical size below which a particle remains in a 
single-domain state during switching.

 Approximating our single domain by an ellipsoid of revolution, the critical 
radius $R_{sd} $ (minor axis, $a=b$) for a single-domain particle can be 
expressed~\cite{Frei}, and calculated by comparing the exchange energy averaged over the 
ellipsoid volume to the magnetostatic energy; [Appendix A]. One finds:

\[R_{sd} =\sqrt{\frac{6A}{N_{C} M_{S}^{2} } [\ln (\frac{4R_{sd} }{a_{1} } )-1]} \]

 Where $N_{C} $ is the demagnetization factor along the c axis (Fig. 1) and 
$a_{1}$ is the nearest-neighbor spacing. The critical radius for a single-domain 
particle is dependent on the material parameters $A, a_{1}, M_{S} $.

 If we consider the infinite cylindre case $(R_{sd} \to\infty , N_{C} \to 0)$, 
in the single domain case there is a critical size between the two 
magnetization reversal processes. When the magnetic easy axis is aligned with 
the applied field, the critical radius $r_{c} =\frac{dc}{2} $ for the 
transition between coherent and curling effect is defined by~\cite{Aharoni1}:

\[r_{c} =\sqrt{\frac{2A}{N_{a} } }  \frac{q}{M_{s} } \] 

where $q=1.8412$ for a cylinder is the first maximum of the Bessel equation 
$\frac{d}{d_{x} } (J_{1} (q))=0$, where $J_{1} (q)$ is the ordinary. Bessel's 
function of the first kind and $N_{a} =2\pi $ is the demagnetizing factor along 
the minor axis of the infinitely long cylinder. Hence:

\[r_{c} =\sqrt{\frac{A}{\pi } }  \frac{q}{M_{S} } \] 

Coherent and curling model rotation are illustrated in Fig.8

 \begin{figure}[htbp]
\centerline{\includegraphics[width=4in,height=3.4in]{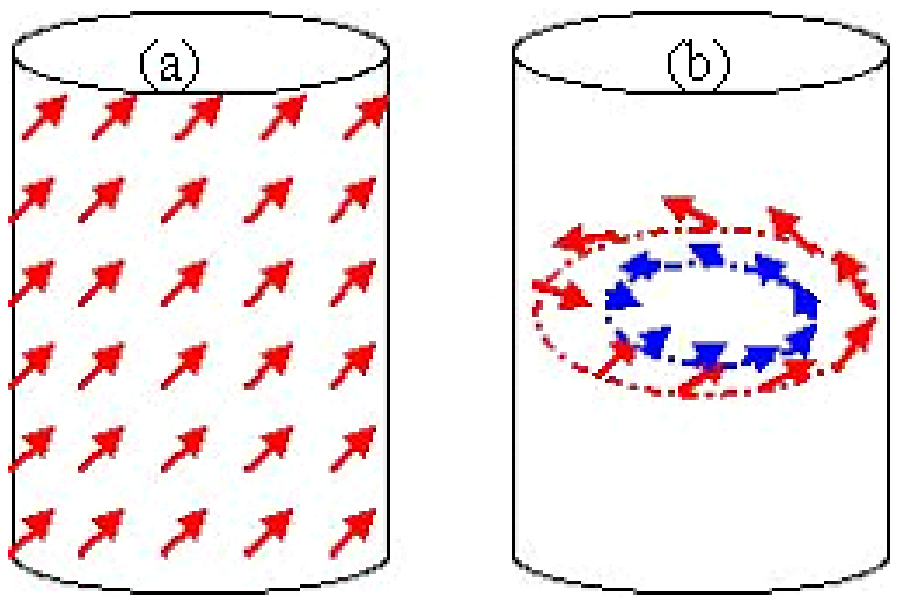}}
  \caption{Illustration of magnetization reversal in a 
single-domain cylinder: (a) coherent rotation and (b) curling case.}
\label{fig8}
\end{figure} 

 For the theoretical explanation of these two magnetization modes, we consider 
the elongated single-domain particle with uniaxial anisotropy shown in Fig.1.

 \subsection{Coherent rotation}

 In the Stoner-Wohlfarth (SW) coherent rotation model~\cite{Stoner2}, the total free energy 
consists of the crystalline and shape anisotropy and Zeeman energy due to the 
external magnetic field.

\begin{equation} 
E=K_{} \sin ^{2} \theta -M_{S} H\cos (\theta -\varphi ),
\end{equation}

Where $(\theta)$ is the angle between the $M_{S} $ and the easy axis, $K$ is the 
effective anisotropy, $(\varphi)$ is the angle between the external applied field and the 
easy direction; $(\theta -\varphi )$ is the angle between $M_{S} $ and $H$. 

\[K_{} =\frac{M_{S}^{2} }{2} (N_{a} -N_{C} )+K_{1} \] 

Here $N_{a} $ and $N_{C} $ are the demagnetizing factor of the ellipsoid along 
the minor and the long axis respectively, and $K_{1} $ is the 
magnetocrystalline anisotropy contant, $K_{1} $ is small by comparing with  
$\frac{M_{S}^{2} }{2} (N_{a} -N_{C} )$.

 After using $\frac{\partial E}{\partial \theta } =0$ and $\frac{\partial ^{2} 
E}{\partial \theta ^{2} } =0$ we obtain: \\ 

 \[H_{S}=\frac{H_{K} }{[\sin \varphi^{\frac{2}{3} } +\cos \varphi ^{\frac{2}{3} }]^{\frac{3}{2} } }\]

 where $H_{S} $ is the switching field and $H_{K} $ is the effective anisotropy 
field : $H_{K} =\frac{2K}{M_{S} } $.

  (The detailed calculation is in appendix B).

 On the other hand, the coherent rotation mode~\cite{Frei} gives the highest and the 
lowest coercivity field $H_{C} (\varphi )$ value of $H$ parallel and 
perpendicular to the easy axis, respectively, for the $15$ and 50 nm of 
diameter.

 In the Stoner-Wohlfarth model the switching field $(H_{S} )$, does not 
represents the coercivity, $H_{C} (\varphi )$, in all cases. However, from the 
discussion in~\cite{Stoner2}, the coercivity can be written as 

\begin{equation} 
H_{C} =\left\{\begin{array}{l} {H_{C1} } \\ {H_{C2} } \end{array}\right.
 for  \left\{\begin{array}{l} {0\le \varphi \le \pi /4,} \\ {\pi /4\le \varphi 
\le \pi /2,} \end{array}\right.
\end{equation}
 with

\[H_{C1} =\left|H_{S} \right|=\frac{2K}{M_{S} } \frac{1}{[\sin \varphi 
^{\frac{2}{3} } +\cos \varphi ^{\frac{2}{3} } ]^{\frac{3}{2} } } \] 

\[H_{C2} =2\left|H_{S} (\varphi =\frac{\pi }{4} )\right|-\left|H_{S} (\frac{\pi 
}{4} \le \varphi \le \frac{\pi }{2} )\right|.\] 

\begin{figure}[htbp]
\centerline{\includegraphics[width=4in,height=3.4in]{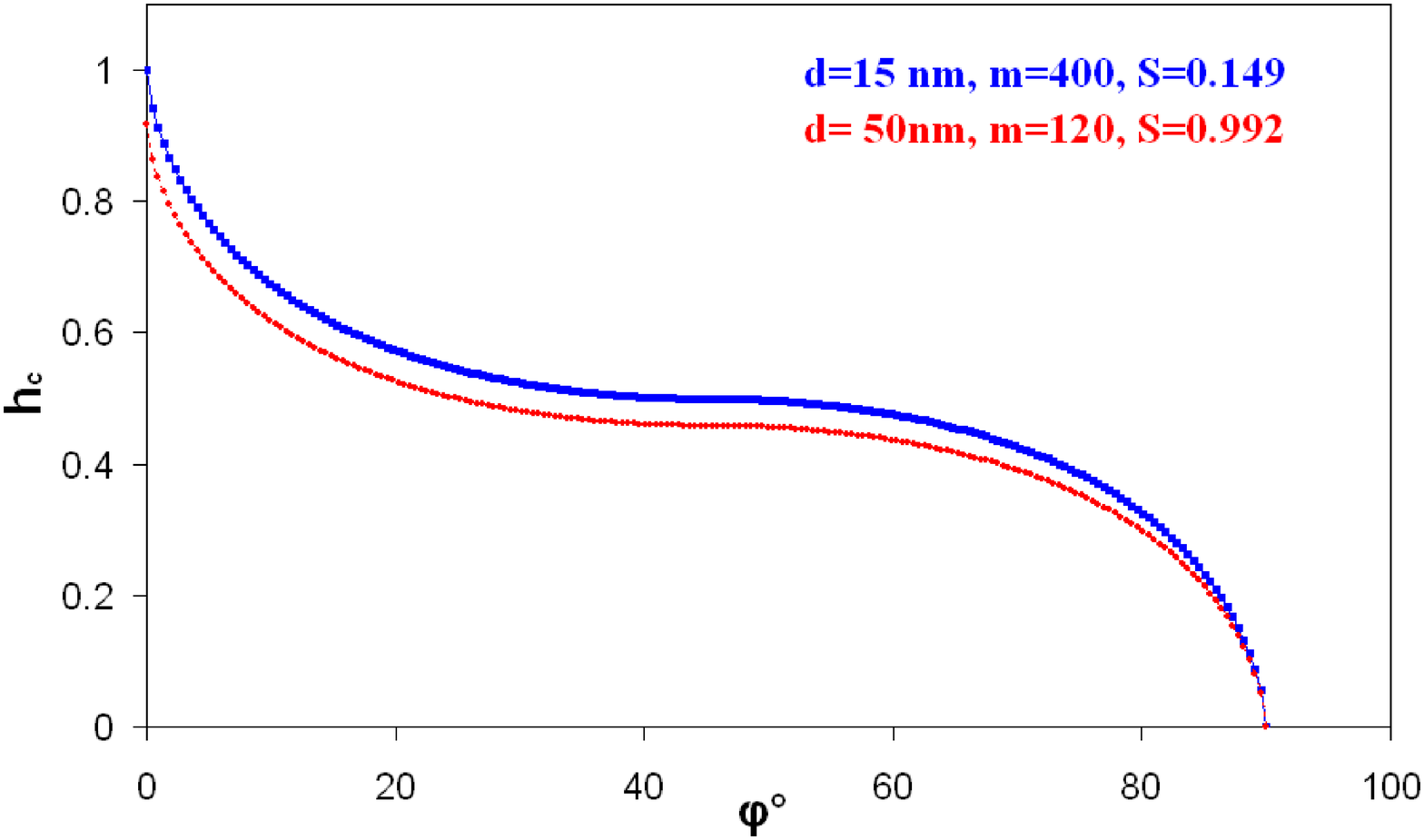}}
  \caption{Angular dependence of the reduced coercivity,
 $h_{C} =\frac{H_{C} }{2\pi M_{S} } $, in Stoner Wolhfarth case, where $\varphi $ is 
the angle between the external applied field and the wire axis, $d$is the wire 
diameter, $m=\frac{c}{a} $is the aspect ratio and $S=\frac{R}{R_{0} } $is the 
reduced ratio.}
\label{fig9}
\end{figure} 

In this mode, two regimes can be identifed, when the applied field is close to 
the magnetic easy axis, $0{}^\circ \le \varphi \le 45{}^\circ $, the hysteresis 
loop is relatively square; hence, the change in sign of the magnetization, 
corresponding to the coercivity, occurs at the switching field.

 In this regime the coercivity field $H_{C1} $ is equal to the switching field 
$H_{S} $.

 For $45{}^\circ \le \varphi \le 90{}^\circ $ , the applied field is oriented 
closer to the magnetic hard axis, and the hysteresis loop is sheared such that 
switching occurs after the magnetization changes sign. In this case, $H_{C} \ne 
H_{S} $ and the coercive field is determined from equation $(H_{C2} )$.

 \subsection{Curling}

 The magnetization curling mode was defined by Frei et al~\cite{Frei} and after it has 
been used for different structures to investigate the magnetic switching of 
films~\cite{Ishii1}, spherical particles, prolate ellipsoids~\cite{Frei} and cylinders~\cite{Ishii2}.  For 
particle size larger than the critical size but still in the single-domain 
regime, magnetization reversal occurs by curling~\cite{Frei, Ishii1}. In the curling model, 
magnetization switching is an abrupt process, and the switching field is very 
close to the nucleation field; hence, $H_{C} =H_{S} $ for all angles.

 The curling nucleation field is given by~\cite{Aharoni2}.

\begin{equation} 
 h\cos (\varphi -\theta )=\frac{1}{2\pi } (N_{a} \sin ^{2} \theta +N_{C} \cos 
^{2} \theta )-\frac{k}{S^{2} } ,
\end{equation}

 where $S$ is the reduced radius defined as $R/R_{0} $, $R$ is the radius of 
the cylinder and R0 is the exchange length defined by $R_{0} =\frac{\sqrt{A} 
}{M_{S} } $, and $R_{0} =25.2 $ nm for Ni. Here $h=\frac{H}{2\pi M_{S} } $ is 
the reduced field~\cite{Ishii2} and the parameter $k=1.08$, which is $q^{2} /\pi $ in the 
notation of Ref.~\cite{Aharoni1}, is a monotonically decreasing function of the aspect ratio 
$m=\frac{c}{a} $ of the ellipsoid (Fig. 1).

 To complete the calculation it is still necessary to eliminate the angle 
$\theta $ from \textbf{(3).}

 By using the fact that it is the same angle as in: 

\begin{equation} 
  h\sin (\varphi -\theta )=\frac{1}{2\pi } (N_{a} -N_{C} )\sin 2\theta 
\end{equation}

 it will become after the differentiating $E$ with respect to $\theta $.

\[\frac{\partial E_{} }{\partial \theta } =2K_{} \sin \theta \cos \theta +M_{S} 
H\sin (\theta -\varphi )=0.\] 
One way of doing it to consider Eqs \textbf{(3)} and \textbf{(4)} as two linear 
equations in $h\sin \varphi $ and $h\cos \varphi $ and to solve them as such 
obtaining:

\begin{eqnarray} 
h\sin \varphi =(\frac{1}{2\pi } N_{a} -\frac{k}{S^{2} } )\sin \theta  \nonumber \\
h\cos \varphi =(\frac{1}{2\pi } N_{C} -\frac{k}{S^{2} } )\cos \theta 
\end{eqnarray}

Adding the squares of these equations by using:

\[\cos ^{2} \theta +\sin ^{2} \theta =1\] 
leads to 

\[ h_{S} =\frac{(\frac{N_{C} }{2\pi } -\frac{k}{S^{2} } )(\frac{N_{a} }{2\pi } -\frac{k}{S^{2} } )}{\sqrt{(\frac{N_{C} }{2\pi } - \frac{k}{S^{2} } )^{2} \sin ^{2}\varphi +(\frac{N_{a} }{2\pi } - \frac{k}{S^{2} } )^{2} \cos ^{2} \varphi } } \]

 where $h_{S} =\frac{H_{S} }{2\pi M_{S} } $.

 For the infinite cylinder, $N_{C} =0, N_{a} =2\pi $, and $k=1.08$ we obtain

\[h_{S} =h_{C} =\frac{1.08 (1-1.08S^{-2} )}{S^{2}  [1.1664S^{-4} +\cos ^{2} 
\varphi  (1-2.16S^{-2} )]^{\frac{1}{2} } } .\] 

 \begin{figure}[htbp]
\centerline{\includegraphics[width=4in,height=3.4in]{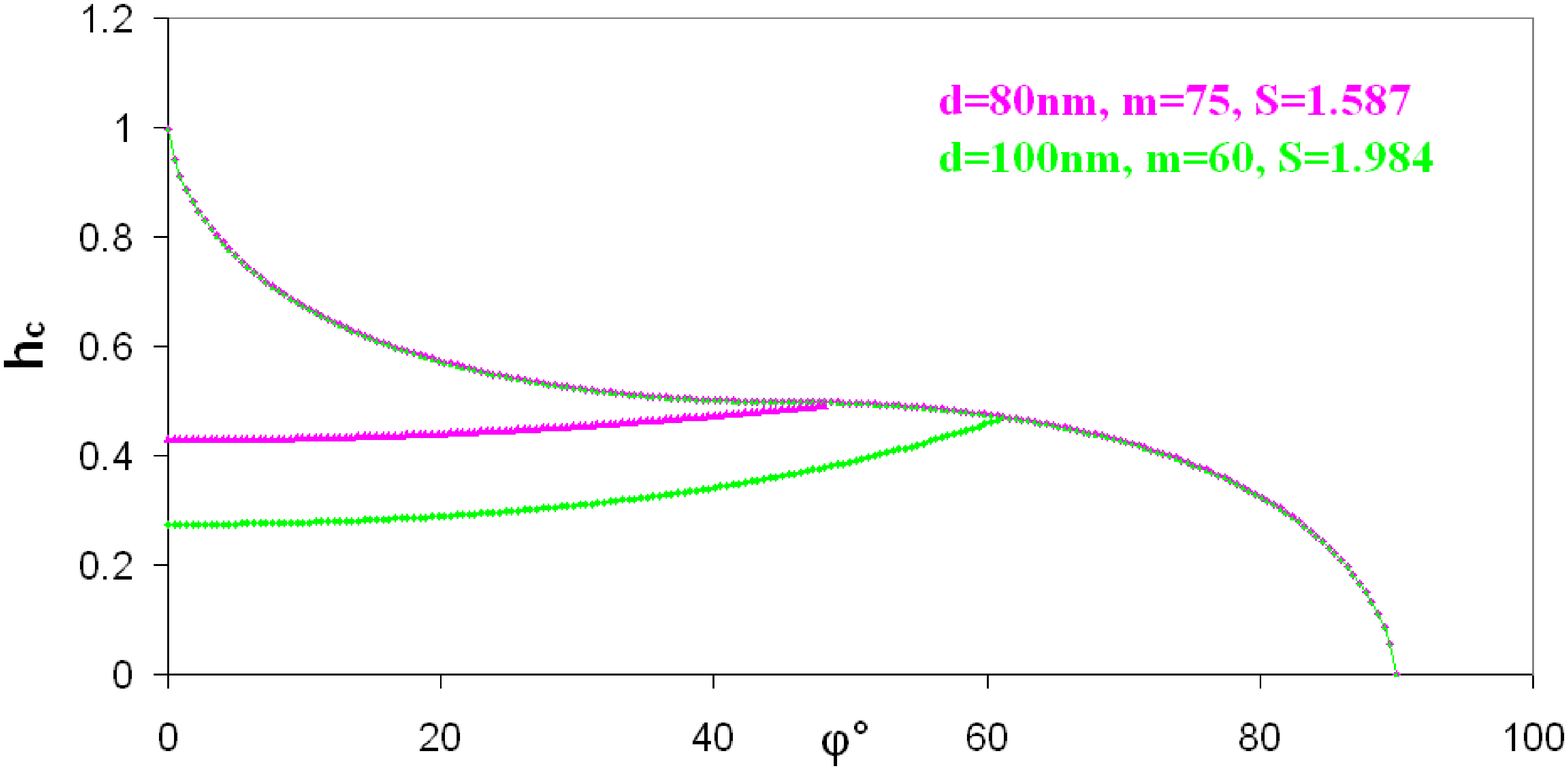}}
  \caption{Angular dependence of the reduced coercivity, $h_{C} 
=\frac{H_{C} }{2\pi M_{S} } $, in the Curling case, where $\varphi $ is the 
angle between the external applied field and the wire axis, $d$is the wire 
diameter, $m=\frac{c}{a} $is the aspect ratio and $S=\frac{R}{R_{0} } $is the 
reduced ratio.}
\label{fig10}
\end{figure}

 The coercivities of infinite cylinders with shape anisotropy depends only on 
the value of $S$ and the angle $\varphi $ between the easy axis of 
magnetization and the direction of measurment as shown in (Fig. 9-10) and 
plotted in reduced units $h_{c} (\varphi )$ $(0<h_{c} <1)$ for various values 
of $S$. $H_{C} $ versus $\varphi $ curves are illustrated in Fig. 11. When we 
compare the curves in Fig. 7a and Fig. 11, the coherent magnetization 
reversal is for the 15 nm and 50 nm of diameters and the curling mode is 
for the 80 nm and 100 nm of diameters. 

 \begin{figure}[htbp]
\centerline{\includegraphics[width=4in,height=3.4in]{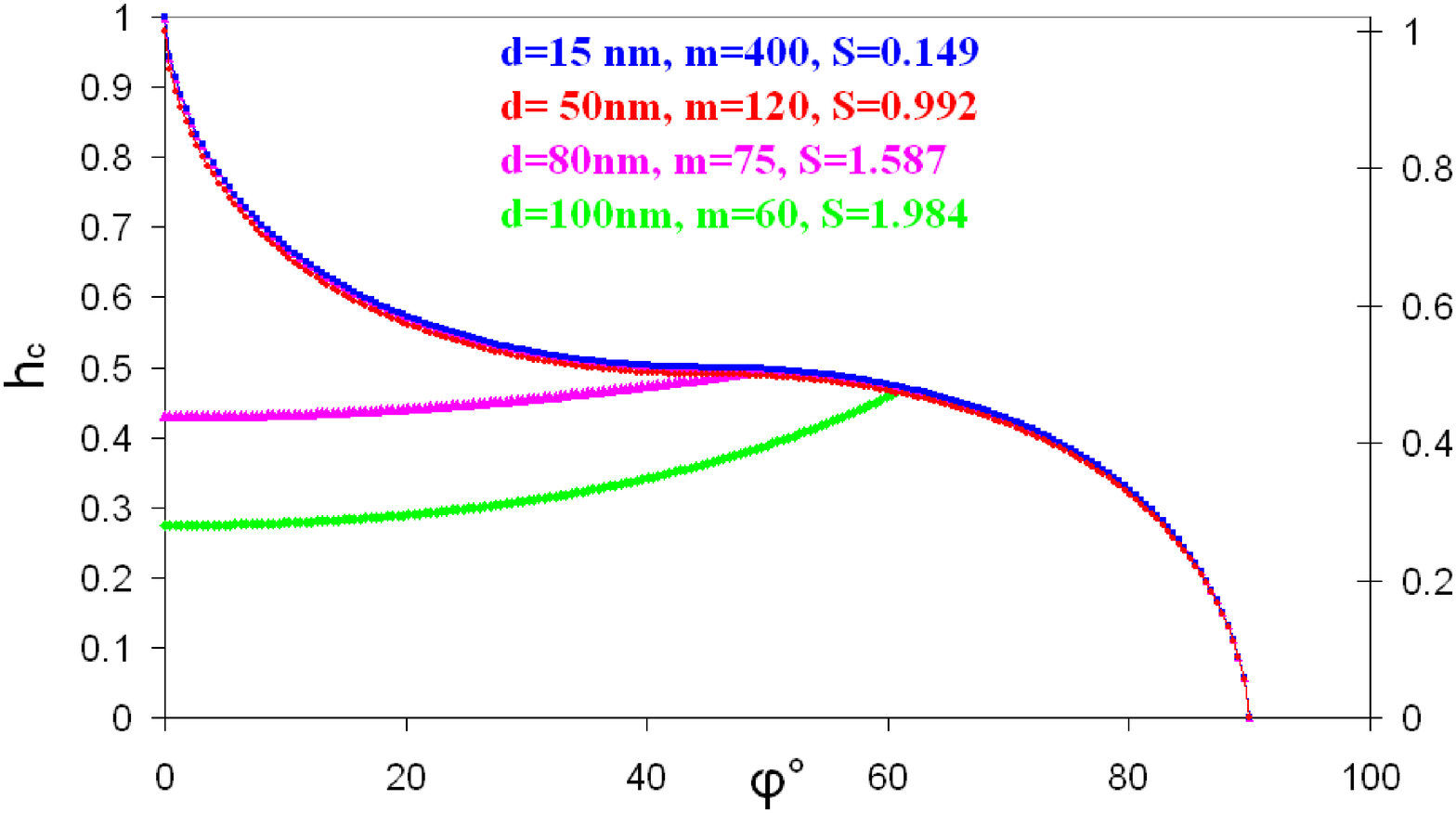}}
  \caption{ Reduced coercivity $h_{c} $ for aligned array of infinitely 
long single-domain cylindrical nanowires as function of angle $\varphi $ 
between easy axis (i.e. cylindrical axis) and direction of measurement, where 
$\varphi $ is the angle between the external applied field and the wire axis, 
$d$is the wire diameter, $m=\frac{c}{a} $is the aspect ratio and 
$S=\frac{R}{R_{0} } $is the reduced ratio.}
\label{fig11}
\end{figure} 

In order to observe small variation in the coercivity, very well geometrically 
characterized samples need to be measured with well controlled inter-wire interaction. 
In our samples the center 
to center distance is fixed, and the strength of interactions is different from 
sample to sample, making direct comparison difficult. The fitted values for the 
coercivity are larger than the experimental data and $H_{C} (\varphi )$ values 
are actually much smaller than $2\pi M_{S} =3050$ Oe, which is expected for 
an individual wire.

 We ascribe such differences between calculations and experimental results to 
the interaction of each wire with the stray field produced by the array. This 
field originating from the effective ferromagnetic coupling between neighboring 
wires reduces the coercive field~\cite{Hertel, Bahiana}.

 Fig. 6 a, shows there is a flat behaviour in the $H_{C} (\varphi )$ curve at 
low angles for 80 and 100 nm of diameters, which may be caused by the 
variation of wire orientation with respect to the normal to the membrane plane.

 These results suggest that the interwire coupling cannot be neglected even for 
PCTE membranes.

 \section{Discussion}

 In the case of a single domain wire, we have to consider three different field 
acting on the wire~\cite{Han}:

\[H_{eff} =H_{Sh} +H_{A} +H_{D} \] 

where $H_{Sh} $ is the demagnetization field (shape), $H_{A} $ is the effective 
anisotropy field and $H_{D} $ is the average dipolar field between the wires.

The demagnetization field (shape) $H_{Sh} $ of the individual wire if magnetized 
parallel to the pore axis is of the order of $2\pi M_{S} =3050 $ Oe with $M_{S} 
$ equal to $485$ emu/cm$^{3} $.

The effective anisotropy field $H_{A} $, given by $H_{A} =-\frac{4}{3} 
\frac{K_{1} }{M_{S} } =123 $ Oe with $K_{1} =-4.5\times 10^{4} $ erg/cm$^{3}$ is 
the cubic magnetocrystalline anisotropy~\cite{Chikazumi} of Ni.

The in plane average dipolar field between the wires, which tends to induce a 
magnetic easy axis perpendicular to the wire axis.

 For a qualitative analysis, we consider a two-dimensional infinite array of 
magnetic dipoles. The array is formed by a square dipole mesh of size $D$. When 
we assume all the moments are aligned along the wires, the average dipolar 
field, which acts on each wire, is parallel to the axis of the wire and may be 
expressed as~\cite{Strijkers, Jackson}:

 \begin{equation} 
H_{D_{0} } =[4.2M_{S} (\pi d^{2} /4)L]/D{}^{3} 
\end{equation}

 On the other hand, for the same array with magnetic moments aligned 
perpendicularly to the axes of the wires, the average dipolar field which acts 
on every wire may be expressed as~\cite{Strijkers, Jackson}:

\begin{equation} 
H_{D_{90} } =[-2.1M_{S} (\pi d^{2} /4)L]/D{}^{3} 
\end{equation}

 where $d$ is the diameter of the wire, $L$ is the length, and $D$ is the 
inter-wire distance [Appendix C]. 

 While the total effective fields following the two directions parallel 
$(0{}^\circ )$ and perpendicular $(90{}^\circ )$ to the wire axis may be 
written:

\[H_{eff} (total)=\left\{_{\begin{array}{l} {} \\ {(90{}^\circ) \to H_{D_{90} } 
+H_{Sh} } \end{array}}^{\begin{array}{l} {(0{}^\circ) \to H_{D_{0} } +H_{A} } \\ 
{} \end{array}} \right. \]

 The calculation for the various angles, show that $H_{C} (\varphi )$ decreases 
as the angle increases, with a bell-type variation. Therefore, the 
magnetization reversal was determined mainly by the curling mode, as shown in 
(Fig. 6a) for 80 nm and 100 nm diameters. For Ni nanowires, the curling is 
predicted for $d\ge 50$ nm, this value agrees with prior work on ordered Ni 
nanowire arrays in alumina~\cite{Neilsch} where the critical diameter was 40 nm. The 
coherent rotation will occur for $15$ and 50 nm wires, which is consistent 
with results in (Fig. 6a). In the same way, the squareness $SQ(\varphi )$ of 
hysteresis loops is measured for different orientations of wires as a function 
of the applied magnetic field. It can be seen that for large wire diameters the 
maximum $SQ(\varphi )$ is found when the magnetic field is applied parallel to 
the axis of the wire and the minimum occurs when the applied field is 
perpendicular to the axis of the wire (see Fig. 6.b). This concurs with the 
abovementioned idea that for the larger wire diameters, the easy axis of 
magnetization is parallel to the axis of the wires. On the other hand, for the 
50 nm of diameters of wires (see Fig. 6.b), the behavior of $SQ(\varphi )$ 
indicates that we are in the crossover region where the axes begin to rotate 
from parallel to perpendicular to the wire axis. But in the wires of 15 nm of 
diameter, the behavior of $SQ(\varphi )$ indicates that we are in the 
perpendicular to the axes of the wires becomes the dominant easy axis of 
effective anisotropy.

 \section{Conclusion}

 In conclusion, by means of theoretical studies and experimental measurements, 
we have investigated the reversal processes in ferromagnetic nanowires. Our 
systematic studies of the effect of the nanowire size diameter show that the 
magnetization reversal mechanism is strongly influenced by the diameter value 
of the nanowires. The various types of $H_{C} (\varphi )$ curves observed in 
(Fig. 7) suggest that no simple magnetization reversal mode could account for 
the complex coercivity mechanism. Two reversal modes are considered as the most 
important: coherent rotation and curling. Good agreement between the measured 
magnetic properties of Ni nanowires and the theoretical calculations is 
obtained. However, further experimental work remains to be done in order to 
observe this transition. \\

{\bf Acknowledgments:} \\

 The authors wish to acknowledge Prof. C. Tannous (LMB) for his valuable 
discussions and help. \\


 \section{Appendix A}

 {\bf Critical radius $R_{sd}$ for single domain} \\

 We consider in Fig. 1, $a=b=R$ as the short axis, and $c$ as the long axis of 
the ellipsoid

\[E_{X} =A[(\frac{d\theta }{dr} )^{2} +\frac{1}{r^{2} } \sin ^{2} \theta ]\] 
Approximating:    $\theta (r)\to \frac{\pi }{2} $
 The average over ellipsoid:  \[\bar{E}_{X} =\frac{1} {\frac{4}{3} \pi R^{2} c} 
\int \frac{A}{r^{2} } 2\pi r dr dZ \]

 with equation:   $\frac{r^{2} }{R^{2} } +\frac{Z^{2} }{c^{2} } =1 \Rightarrow 
Z=\pm c\sqrt{1-\frac{r^{2} }{R^{2} } } $

\[\Rightarrow \bar{E}_{X} =\frac{1}{{\raise0.7ex\hbox{$ 4 $}\!\mathord{\left/ 
{\vphantom {4 3}} \right. \kern-\nulldelimiterspace}\!\lower0.7ex\hbox{$ 3 $}} 
\pi R^{2} c} \int _{{\raise0.7ex\hbox{$ a_{1}  $}\!\mathord{\left/ {\vphantom 
{a_{1}  2}} \right. \kern-\nulldelimiterspace}\!\lower0.7ex\hbox{$ 2 $}} 
}^{R}2\pi A\frac{dr}{r} \int _{0}^{c\sqrt{(1-\frac{r^{2} }{R^{2} } )} }2dZ  \] 
where $a_{1} $ is the nearest-neighbor spacing

\[\Rightarrow \bar{E}_{X} =\frac{3A}{R^{2} c} \int _{{\raise0.7ex\hbox{$ a_{1}  
$}\!\mathord{\left/ {\vphantom {a_{1}  2}} \right. 
\kern-\nulldelimiterspace}\!\lower0.7ex\hbox{$ 2 $}} }^{R}\frac{dr}{r} \int 
_{0}^{c\sqrt{(1-\frac{r^{2} }{R^{2} } )} }dZ  \] 

\[\Rightarrow \bar{E}_{X} =\frac{3A}{R^{2} c} \int _{{\raise0.7ex\hbox{$ a_{1}  
$}\!\mathord{\left/ {\vphantom {a_{1}  2}} \right. 
\kern-\nulldelimiterspace}\!\lower0.7ex\hbox{$ 2 $}} }^{R}\frac{dr}{r} 
c\sqrt{1-\frac{r^{2} }{R^{2} } }  \] 
changing:   $\frac{r}{R} =\cos x\Rightarrow \frac{dr}{R} =-\sin xdx$

\[\Rightarrow \bar{E}_{X} =\frac{3A}{R^{2} } \int _{0}^{\arccos 
({\raise0.7ex\hbox{$ a_{1}  $}\!\mathord{\left/ {\vphantom {a_{1}  2R}} \right. 
\kern-\nulldelimiterspace}\!\lower0.7ex\hbox{$ 2R $}} )}\frac{R\sin ^{2} 
x}{R\cos x} dx \]

\[\Rightarrow \bar{E}_{X} =\frac{3A}{R^{2} } \int _{0}^{\arccos 
({\raise0.7ex\hbox{$ a_{1}  $}\!\mathord{\left/ {\vphantom {a_{1}  2R}} \right. 
\kern-\nulldelimiterspace}\!\lower0.7ex\hbox{$ 2R $}} )}\frac{\sin ^{2} x}{\cos 
x} dx \] 
From the integral:   $\int \frac{\sin ^{2} x}{\cos x} dx=-[\sin x+\ln 
tg(\frac{\pi }{4} -\frac{x}{2} )] $

 Taking:   $u=\arccos ({\raise0.7ex\hbox{$ a_{1}  $}\!\mathord{\left/ 
{\vphantom {a_{1}  2R}} \right. \kern-\nulldelimiterspace}\!\lower0.7ex\hbox{$ 
2R $}} )\Rightarrow {\raise0.7ex\hbox{$ a_{1}  $}\!\mathord{\left/ {\vphantom 
{a_{1}  2R}} \right. \kern-\nulldelimiterspace}\!\lower0.7ex\hbox{$ 2R $}} 
=\cos u$

\[\Rightarrow \bar{E}_{X} =-\frac{3A}{R^{2} } [\sin u+\ln tg(\frac{\pi }{4} 
-\frac{u}{2} )]\]

 using:   $B=\ln tg(\frac{\pi }{4} -\frac{u}{2} )=\ln \left[\frac{\cos 
\frac{u}{2} -\sin \frac{u}{2} } {\cos \frac{u}{2} +\sin \frac{u}{2} } \right]$

 with   $t=tg\frac{u}{2} ,$ $B=\frac{1-t}{1+t} =\frac{(1-t)^{2} }{1-t^{2} } 
=\frac{1+t^{2} -2t}{1-t^{2} } $

\[\Rightarrow B=\frac{1}{\cos u} -tgu=\frac{1-\sin u}{\cos u} 
=\frac{1-\sqrt{1-(\frac{a_{1} }{2R} )^{2} } }{\frac{a_{1} }{2R} } \] 

\[\Rightarrow B=\frac{1-(1-\frac{1}{2} (\frac{a_{1} }{2R} )^{2} 
.....)}{\frac{a_{1} }{2R} } ,\Rightarrow B\cong \frac{a_{1} }{4R} \] 

\[\Rightarrow \bar{E}_{X} =-\frac{3A}{R^{2} } \left[1+\ln \frac{a_{1} }{4R} 
\right]=\frac{3A}{R^{2} } \left[\ln \frac{4R}{a_{1} } -1\right]\] 
Comparing this energy to the magnetostatic energy:

 We get:     $E_{D} =\frac{1}{2} M_{S}^{2} N_{c} $

\[\Rightarrow R_{sd} =\sqrt{\frac{6A}{N_{C} M_{S}^{2} } [\ln (\frac{4R_{sd} 
}{a_{1} } )-1]} \]

 \section{Appendix B}

{\bf Stoner-Wohlfarth (SW) coherent rotation model} \\

 The total free energy, (see Fig. 1)

\[E=K_{} \sin ^{2} \theta -M_{S} H\cos (\theta -\varphi )\]

 After differentiating $E$ with respect to $\theta $, we obtain 

\[\frac{\partial E_{} }{\partial \theta } =2K_{} \sin \theta \cos \theta +M_{S} 
H\sin (\theta -\varphi ),\] 
We find that for a minimum in $E$, $\frac{\partial E_{} }{\partial \theta } 
=0$, i. e;

\[2K_{} \sin \theta \cos \theta +HM_{S} \sin (\theta -\varphi )=0\] 
After using the trigonometry equation:

$\sin (\theta -\varphi )=\sin \theta \cos \varphi -\cos \theta \sin \varphi $ 
We find:

\[[2K_{} +\frac{HM_{S} \cos \varphi }{\cos \theta } -\frac{HM_{S} \sin \varphi 
}{\sin \theta } ]\sin \theta \cos \theta =0\] 
Substituting: $\begin{array}{l} {H_{Z} =H\cos \varphi } \\ {H_{X} =H\sin 
\varphi } \end{array}$

 If we exlude the extrema at $\theta =0, \pi $ and $\theta =\frac{\pi }{2} 
, \frac{3\pi }{2} $, we can assume that above terms (in square brackets) is zero.

\[2K+\frac{H_{z} M_{S} }{\cos \theta } -\frac{H_{x} M_{S} }{\sin \theta } =0\]

 Defining:    $\begin{array}{l} {h_{z} =\frac{H_{Z} M_{S} }{2K} =\frac{H_{Z} 
}{H_{K} } } \\ {h_{x} =\frac{H_{X} M_{S} }{2K} =\frac{H_{X} }{H_{K} } } 
\end{array}$,  $H_{K} =\frac{2K}{M_{S} } $

 The second derivation gives: \[\frac{\partial ^{2} E}{\partial \theta ^{2} } 
=\frac{h_{x} \sin \theta }{\cos ^{2} \theta } +\frac{h_{z} \cos \theta }{\sin 
^{2} \theta } \]

 After using $\frac{\partial E}{\partial \theta } =0$ and $\frac{\partial ^{2} 
E}{\partial \theta ^{2} } =0$, we obtain:

\[\begin{array}{l} {\frac{h_{x} }{\cos \theta } -\frac{h_{z} }{\sin \theta } =1 
      (B.1)} \\ {\frac{h_{z} }{\sin ^{3} \theta } =-\frac{h_{x} }{\cos ^{3} 
\theta }       (B.2)} \end{array}\]

 Eliminating $\theta $ from these equation we obtain:

 By (B.2),    $h_{x} =- \frac{\sin ^{3} \theta }{\cos ^{3} \theta }  h_{z} $

 Trough (B.1),   $- \frac{\sin ^{2} \theta }{\cos ^{3} \theta }  h_{z} 
-\frac{h_{z} }{\cos \theta } =1$

 Factorizing: $\frac{-1}{\cos ^{3} \theta }  [\sin ^{2} \theta +\cos ^{2} \theta 
]  h_{z} =1$

 We get  $\left\{\begin{array}{l} {h_{z} =-\cos ^{3} \theta } \\ {h_{x} =\sin 
^{3} \theta } \end{array}\right. ,$

 which gives us  $h_{x}^{{\raise0.7ex\hbox{$ 2 $}\!\mathord{\left/ {\vphantom 
{2 3}} \right. \kern-\nulldelimiterspace}\!\lower0.7ex\hbox{$ 3 $}} } 
+h_{z}^{{\raise0.7ex\hbox{$ 2 $}\!\mathord{\left/ {\vphantom {2 3}} \right. 
\kern-\nulldelimiterspace}\!\lower0.7ex\hbox{$ 3 $}} } =1$,  (Astroid equation)

 Replacing $h_{x} $ and $h_{z} $ by these values we find the switching field 
$H_{S} $ :

\[H_{S} =\frac{H_{K} }{[\sin \varphi ^{\frac{2}{3} } +\cos \varphi 
^{\frac{2}{3} } ]^{\frac{3}{2} } } \]

 \section{Appendix C}

{\bf Dipolar interaction field} \\

 If we have a single dipole $\vec{P}_{0} $ surrounded by an array of dipoles 
$\vec{P}_{i} $ (Fig. C1), the interaction is:

\[E=\sum _{i}\frac{\vec{P}_{0} .\vec{P}_{i} }{r^{3} }  - \frac{3(\vec{P}_{0} 
.\vec{r})(\vec{P}.\vec{r})}{r^{5} } \] 
This is equivalent to a field $H_{D} $ acting on the dipole such that: 
$E=-\vec{P}_{0} .\vec{H}_{D} $

\[\Rightarrow -\vec{P}_{0} .\vec{H}_{D} =\sum _{i}\frac{\vec{P}_{0} 
.\vec{P}_{i} }{r^{3} }  - \frac{3(\vec{P}_{0} .\vec{r})(\vec{P}.\vec{r})}{r^{5} 
} \] 

 \begin{figure}[htbp]
\centerline{\includegraphics[width=4in,height=3.4in]{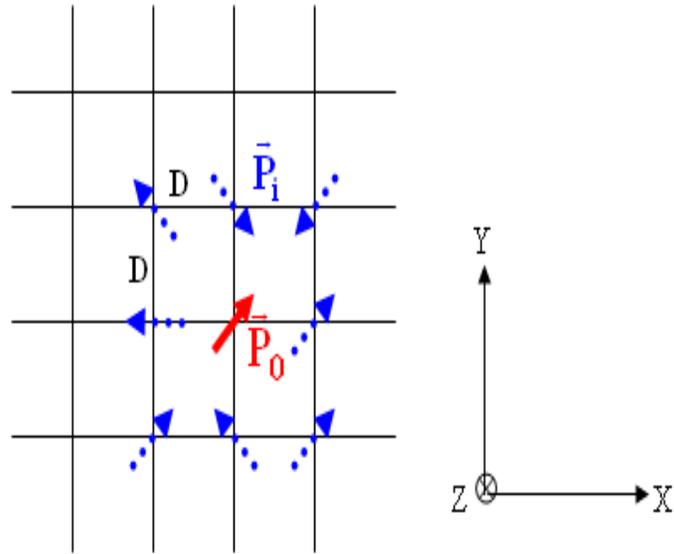}}
  \caption{ Dipole $\vec{P}_{0} $ surrounded by a number of $\vec{P}_{i}$ 
dipoles sitting on a $2D$ square lattice with $D$  the average dipole 
separation.
}
\label{figC1}
\end{figure}  

\textbf{Calculating }$H_{D_{Z} } $

 Assuming the dipoles $\vec{P}_{i} $ sit on a $2D$ square lattice of parameter 
$D$ (the average nanowire separation)

 $\left|\vec{P}_{0} \right|=1$, is directed along the $Z$ axis,

 hence  $\vec{P}_{0} =\left[\begin{array}{l} {0} \\ {0} \\ {1} 
\end{array}\right. $    $\Rightarrow $ $-\vec{P}_{0} .\vec{H}_{D} =-H_{D_{Z} } $

 and  $\vec{P}_{i} =\left[\begin{array}{l} {P_{x} } \\ {P_{y} } \\ {P_{Z} } 
\end{array}\right. $,  and $\vec{r}=\left[\begin{array}{l} {iD} \\ {jD} \\ {0} 
\end{array}\right. $  $\Rightarrow $  $\vec{P}_{0} .\vec{r}=0,$ $\vec{P}_{i} 
.\vec{P}=P_{Z} $, 

 we have  $r^{3} =D^{3} (i^{2} +j^{2} )^{{\raise0.7ex\hbox{$ 3 
$}\!\mathord{\left/ {\vphantom {3 2}} \right. 
\kern-\nulldelimiterspace}\!\lower0.7ex\hbox{$ 2 $}} } $

 Then,  $H_{D_{Z} } =-\sum \frac{P_{Z} }{r^{3} }  $

 $P_{Z} =M_{S} V$(All surrounded dipoles saturated along $Z$):

\[\Rightarrow H_{D_{Z} } =-\frac{M_{S} V}{D^{3} } \sum _{i,j}\frac{1}{(i^{2} 
+j^{2} )^{{\raise0.7ex\hbox{$ 3 $}\!\mathord{\left/ {\vphantom {3 2}} \right. 
\kern-\nulldelimiterspace}\!\lower0.7ex\hbox{$ 2 $}} } }  \] 

\[\sum _{\begin{array}{l} {i,j=-\infty } \\ {(i,j\ne 0)} \end{array}}^{+\infty 
}\frac{1}{(i^{2} +j^{2} )^{{\raise0.7ex\hbox{$ 3 $}\!\mathord{\left/ {\vphantom 
{3 2}} \right. \kern-\nulldelimiterspace}\!\lower0.7ex\hbox{$ 2 $}} } }  =4\sum 
_{\begin{array}{l} {i=1,\infty } \\ {j=1,\infty } \end{array}}\frac{1}{(i^{2} 
+j^{2} )^{{\raise0.7ex\hbox{$ 3 $}\!\mathord{\left/ {\vphantom {3 2}} \right. 
\kern-\nulldelimiterspace}\!\lower0.7ex\hbox{$ 2 $}} } }  \]

 \textit{Numerically:}

\[4\sum _{\begin{array}{l} {i=1,\infty } \\ {j=1,\infty } 
\end{array}}\frac{1}{(i^{2} +j^{2} )^{{\raise0.7ex\hbox{$ 3 $}\!\mathord{\left/ 
{\vphantom {3 2}} \right. \kern-\nulldelimiterspace}\!\lower0.7ex\hbox{$ 2 $}} 
} }  =4\sum _{\begin{array}{l} {i=1,100} \\ {j=1,100} 
\end{array}}\frac{1}{(i^{2} +j^{2} )^{{\raise0.7ex\hbox{$ 3 $}\!\mathord{\left/ 
{\vphantom {3 2}} \right. \kern-\nulldelimiterspace}\!\lower0.7ex\hbox{$ 2 $}} 
} }  \cong 4\times (1.05)\] 

\[\Rightarrow H_{D_{Z} } =\frac{-4.2M_{S} V}{D^{3} } \] 

\textbf{Calculating $H_{D_{X} } $ :}

 Taking,

 $\vec{P}_{0} =\left[\begin{array}{l} {1} \\ {0} \\ {0} \end{array}\right. $, 
 $\vec{P}=\left[\begin{array}{l} {P_{X} } \\ {P_{Y} } \\ {P_{Z} } 
\end{array}\right. $, and  $\vec{r}=\left[\begin{array}{l} {iD} \\ {jD} \\ {0} 
\end{array}\right. $,

\[\Rightarrow E=-\vec{P}_{0} .\vec{H}_{D} =-H_{D_{X} } ,  \vec{P}_{0} 
.\vec{P}=P_{X} \] 

\[\Rightarrow E=\sum _{i}\frac{P_{X} }{r^{3} } -\frac{3(iD)(P_{X} iD+P_{Y} 
jD)}{r^{5} }  \] 
We get $E=\sum _{i}\frac{r^{2} P_{X} -3D^{2} (i^{2} P_{X} +ijP_{Y} )}{r^{5} }  $

\[\Rightarrow H_{D_{X} } =-\sum _{i,j=-\infty }^{+\infty }\frac{P_{X} (i^{2} 
+j^{2} )-3(i^{2} P_{X} +ijP_{Y} )}{D^{3} (i^{2} +j^{2} )^{{\raise0.7ex\hbox{$ 5 
$}\!\mathord{\left/ {\vphantom {5 2}} \right. 
\kern-\nulldelimiterspace}\!\lower0.7ex\hbox{$ 2 $}} } }  \] 
If all the surrounding dipoles are saturated along $X$:

\[\Rightarrow P_{X} =M_{S} V,  P_{Y} =0\]

 we get $H_{D_{X} } =\frac{M_{S} V}{D^{3} } \sum _{i,j=-\infty }^{\infty 
}\frac{(2i^{2} -j^{2} )}{(i^{2} +j^{2} )^{{\raise0.7ex\hbox{$ 5 
$}\!\mathord{\left/ {\vphantom {5 2}} \right. 
\kern-\nulldelimiterspace}\!\lower0.7ex\hbox{$ 2 $}} } }  $$\Rightarrow 
$$H_{D_{X} } =\frac{4M_{S} V}{D^{3} } \sum _{i,j=1}^{\infty }\frac{(2i^{2} 
-j^{2} )}{(i^{2} +j^{2} )^{{\raise0.7ex\hbox{$ 5 $}\!\mathord{\left/ {\vphantom 
{5 2}} \right. \kern-\nulldelimiterspace}\!\lower0.7ex\hbox{$ 2 $}} } }  $

 \textit{Numerically:}

\[\sum _{i,j=1}^{\infty }\frac{(2i^{2} -j^{2} )}{(i^{2} +j^{2} 
)^{{\raise0.7ex\hbox{$ 5 $}\!\mathord{\left/ {\vphantom {5 2}} \right. 
\kern-\nulldelimiterspace}\!\lower0.7ex\hbox{$ 2 $}} } }  \cong \sum 
_{i,j=1}^{100}\frac{(2i^{2} -j^{2} )}{(i^{2} +j^{2} )^{{\raise0.7ex\hbox{$ 5 
$}\!\mathord{\left/ {\vphantom {5 2}} \right. 
\kern-\nulldelimiterspace}\!\lower0.7ex\hbox{$ 2 $}} } }  \cong 0.526\] 

\[\Rightarrow H_{D_{X} } = 4(0.526)\frac{M_{S} V}{D^{3} } =2.1\frac{M_{S} 
V}{D^{3} } \] 

\[\Rightarrow H_{D_{X} } = -\frac{1}{2} H_{D_{Z} } \] 
This shows the dipolar field along x is half the field along $Z$.


\begin{thebibliography}{99}


 \bibitem{Zheng} M. Zheng et al., Phys. Rev. B, \textbf{62}, 12282 (2000).

 \bibitem{Manalis} S. Manalis et al., Appl. Phys. Lett. \textbf{66}, 2585 (1995).

 \bibitem{Yoshida} Yoshida, M. et al., J. Radiation Effects in solids\textbf{ 126}, 409 (1993).

 \bibitem{Piraux} Piraux, L. et al., Appl. Phys. Lett. \textbf{65}, 2484 (1994).

 \bibitem{Brumlik} C. J. Brumlik and C. R. Martin, Analytical Chem. \textbf{59}, 2625 (1992).

 \bibitem{Bertotti} G. Bertotti, {\it "Hysteresis in magnetism",} Academic Press, New-York (1998).

 \bibitem{Aharoni1} A. Aharoni, {\it "Introduction to the theory of ferromagnetism",} Oxford 
University Press, New York, (1996).

 \bibitem{Stoner1} E. C. Stoner and E. P. Wohlfarth, Philos. Trans. R. Soc. London, Ser. A 
\textbf{240}, 599  (1957).

 \bibitem{Skomski} R. Skomski, A. Kashyap, K. D. Sorge, D. J. Sellmyer J. Appl. Phys. 
\textbf{95}, 7022 (2005).

 \bibitem{Kronmuller} H. Kronm\"uller, Phys. Status Solidi B\textbf{ 144}, 385 (1987).

 \bibitem{Sagawa} M. Sagawa, S. Fujimura, N. Tagawa, and Y. Matsuura, J. Appl. Phys. 
\textbf{55}, 2083 (1992).

 \bibitem{Lederman} M. Lederman, R. O'Barr, and S. Schultz, IEEE Trans. Magn. \textbf{31}, 3793 
(1995); R. O'Barr, M. Lederman, S. Schultz, W. H. Xu, A. Scherer, and J. 
Tonucci, J. Appl. Phys. \textbf{79}, 5303 (1996).

\bibitem{Wernsdorfer} W. Wernsdorfer, B. Doudin, D. Mailly, K. Hasselbach, A. Benoit, J. Meier, 
J.-Ph. Ansermet, and B. Barbara, Phys. Rev. Lett. \textbf{77}, 1873 (1996).

\bibitem{Bantu} A. K. M. Bantu, J. Rivas, G. Zaragoza, M. A. Lopez-Quintela, and M. C. 
Blanco, J. Appl. Phys. \textbf{89}, 3393 (2001).

\bibitem{Garcia} J. M. Garcia, A. Asenjo, J. Velazquez, D. Garcia, M. Vazquez, P. Aranda, 
and E. Ruiz-Hitzky, J. Appl. Phys. \textbf{85}, 5480 (1999).

\bibitem{Whitney} T. M. Whitney, J. S. Jiang, P. C. Searson, and C. L. Chien, Science\textbf{ 
261}, 1316  (1993). 

\bibitem{Frei} E.H. Frei, S. Shtrikman, and D. Treves, Phys. Rev. \textbf{106}, 446 (1957).

\bibitem{Stoner2} E. C. Stoner and E. P. Wohlfarth, Phil. Trans. Roy. Soc. Lond. A 
\textbf{240}, 599 (1948). Reprinted in IEEE Trans. Magn. \textbf{27}, 3475 
(1991)

\bibitem{Ishii1} Y. Ishii, S. Hasegawa, M. Saito, Y. Tabayashi, Y. Kasajima, and T. 
Hashimoto, J. Appl. Phys. \textbf{82}, 3593 (1997).

\bibitem{Ishii2} Y. Ishii, M. Sato, J. Appl. Phys. \textbf{65}, 3146 (1989).

\bibitem{Aharoni2} A. Aharoni, J. Appl. Phys.\textbf{ 82,} 1281 (1997).

\bibitem{Hertel} R. Hertel, J. Appl. Phys. \textbf{90}, 5752 (2001).

\bibitem{Bahiana} M. Bahiana, F. S. Amaral, S. Allende, and D. Altbir, Phys. Rev. B. 
\textbf{74}, 174412 (2006).

\bibitem{Han}  G. C. Han, IEEE Trans. Magn. \textbf{38}, No. 5, 2562 (2002).

\bibitem{Chikazumi} S. Chikazumi {\it "Physics of Ferromagnetism",} Wiley, New York 
(1964).

\bibitem{Strijkers} G. J. Strijkers, J. H. J. Dalderop, M. A. Abroeksteeg, H. J. M. Swagten, 
and W. J. M. de Jonge, J. Appl. Phys\textit{.}, \textbf{86}, 5141 (1999).

\bibitem{Jackson} J. Jackson, {\it "Classical Electrodynamics",} Wiley, New York (1962).

\bibitem{Neilsch} K. Neilsch, R. B. Wehrspohn, J. Barthel, J. Kirschner, U. Gl"osele Appl. 
Phys. Lett. \textbf{79}, 9, 1360 (2001).

\end{thebibliography}
\end{document}